\documentclass[twocolumn, times, tighten]{aastex62}
\usepackage{graphicx}
\usepackage{contcaption,mathtools}

\submitjournal{ApJSupplSer}

\shorttitle{Winged radio galaxies} \shortauthors{Bera et al.}

\begin{document}

\title{FIRST winged radio galaxies with {\it X} and {\it Z} symmetry}

\correspondingauthor{Sabyasachi Pal}

\author{Soumen Bera} 
\affil{Department of Physics, Jadavpur University, Kolkata, 700032, India}

\author{Sabyasachi Pal}
\affil{Midnapore City College, Kuturiya, Bhadutala, Paschim Medinipur, West Bengal, 721129, India}
\affil{Indian Centre for Space Physics, 43 Chalantika, Garia Station Road, 700084, India}
\email{sabya.pal@gmail.com}

\author{Tapan K. Sasmal} 
\affil{Department of Physics, Jadavpur University, Kolkata, 700032, India}

\author{Soumen Mondal}
\affil{Department of Physics, Jadavpur University, Kolkata, 700032, India}

\begin{abstract}
{\it X}-shaped radio galaxies are a subclass of radio sources that exhibit a pair of secondary low surface brightness radio lobes oriented at an angle to the primary high surface brightness lobes. Sometimes, the secondary low brightened lobes emerge from the edges of the primary high brightened lobes and form a Z-symmetric morphology. We present a systematical search result for {\it X}-shaped radio galaxies (XRGs) and {\it Z}-shaped radio galaxies (ZRGs) from the VLA Faint Images of the Radio Sky at Twenty-Centimeters (VLA FIRST) Survey at 1.4 GHz. Our search yields a total of 296 number of radio sources, out of which 161 are XRGs and 135 are ZRGs. We have also made optical identification of these sources from the different available literature. J1124+4325 and J1319+0502 are the farthest known XRG and ZRG, respectively. We have estimated spectral index and radio luminosity of these radio sources and made a comparative study with previously detected XRGs and ZRGs. The average value of luminosities for XRGs is higher than that of ZRGs. With the help of a large sample size of the newly discovered XRGs and ZRGs, various statistical properties of these sources are studied. Out of 161 XRGs presented in the current paper, 70\% (113) are FR II radio galaxies and 13\% (20) are FR I radio galaxies. For 28 XRGs, the morphology is complex and could not be classified. For XRGs, the statistical studies are done on the angle between the major axis and minor axis and the relative size of the major and minor axes. For the ZRGs a statistical study is done on the angular size.
\end{abstract}

\keywords{Active galactic nuclei (16); Catalogs (205); Jets (870); Quasars (1319); Radio continuum emission (1340); Surveys (1671)}

\section{Introduction}
\label{sec:intro}
A typical double-lobed radio galaxy morphology shows a pair of radio jets, each one directed in the opposite direction from the central black hole (BH) or active galactic nuclei (AGN). The alignment of the primary jets gives a linear structure in radio map. A small number of subclasses are found with a nonlinear special structure. `Winged' radio galaxies is a subclass where two additional secondary lobes (wings) are found beside the primary lobes. These symmetric and low luminosity extrusion of plasma extend at an angle from the center to a distance nearly equal to or lower than the length of the active lobes. Based on the alignment of the secondary lobes with the primary lobes, two subclasses are defined, {\it `X'}-shaped radio galaxies (XRGs) and {\it `Z/S'}-shaped radio galaxies (ZRGs). For XRGs, both the sets of lobes pass symmetrically through the center of the elliptical galaxy that is the source of the lobes. In the case of ZRGs, the secondary lobes are seen from the edges of the primary jets.

The first reported radio galaxy with wings is 3C 272.1 \citep{Ri72}, which showed a {\it Z}-like structure. Later NGC 326 also showed {\it Z}-like structure with possible precessing beams \citep{Ek78}. The elliptical galaxy NGC 3309 \citep{Dr88} was identified as the {\it X}-shaped source by \citet{Ko90}. \citet{Le92} first classified sources with {\it X}-shapes as XRGs and presented a list of 11 objects with wings. \citet{Ch07} (C07 afterwards) first made a systematic study to look for radio galaxies with wings and identified 100 XRG candidates using the VLA Faint Images of the Radio Sky at Twenty-centimeters (FIRST) survey. Later, (\citet{Ya19} Y19 afterwards) also identified 290 winged radio galaxies from the FIRST survey. \citet{Bh20} discovered 50 XRGs and tens of ZRGs from the TIFR GMRT Sky Survey (TGSS) at 150 MHz. By using an automated morphological classification scheme to the FIRST radio sources, \citet{Pr11} identified 156 XRG candidates out of which 21 sources had already been reported in C07.  

The maximum number of these sources are of Fanaroff-Riley type II (FR II; \citep{Fa74a}) and the remaining are either FR I or mixed \citep{Me02}.

The origin of the wings in XRGs is still an arguable topic. There are several models available to explain the mechanism behind the formation of this exotic type of sources \citep{Ro01, De02}. The proposed models are: (a) the backflow of plasma \citep{Le84, Ca02}, the secondary wings result from backflow of plasma from the hot spots of the active lobes into the surrounding asymmetric medium; (b) a merger of two black holes \citep{Ro01, Me02}: The coalescence of two super-massive black holes (SMBHs) is another possibility for the origin of the {\it X}-shaped radio morphology; (c) realignment of a central SMBH-accretion disk system \citep{De02}, the secondary lobes in XRGs may be the remnants left over from a rapid realignment of a central SMBH-accretion disk system; and (d) precession of twin jets \citep{Ma94}. It is important to note that none of the above-mentioned models can explain the properties of all the XRGs. 

The studies made by C07 and Y19 are far from complete within the FIRST coverage area because both of the studies concentrated on a selected subset of the survey depending on the high dynamic range. C07 also looked for only those sources that are larger than 15$\arcsec$. Naturally many of the {\it X}-shaped sources within the FIRST survey area are missed due to above-mentioned restrictions. We made a complete study to search for all possible XRGs and ZRGs within the FIRST survey area.

In section \ref{sec:source-identification}, we have described source identification method. In section \ref{sec:result}, we described different properties of XRGs (section \ref{subsec:prop-XRG}) and ZRGs (section \ref{subsec:prop-ZRG}). General discussion on different results are made in section \ref{sec:disc} and in section \ref{sec:conclusion}, concluding remarks are made.

We have used the following cosmology parameters for the entire discussion in this paper: $H_0 = 67.4$ km s$^{-1}$ Mpc$^{-1}$, $\Omega_m = 0.315$ and $\Omega_{vac} = 0.685$ \citep{Ag18}. These cosmological constants are estimated using final full-mission Planck measurements of the cosmic microwave background (CMB) anisotropies, combining information from the temperature and polarization maps and the lensing reconstruction.

\section{Identification of the XRGs and ZRGs}
\label{sec:source-identification}

\subsection{The FIRST Survey Data}
\label{subsec:first}
The FIRST survey \citep{Wh97} covers a radio sky of 10,575 square degrees of the north and south Galactic caps near 1400 MHz (21 cm). This survey has a typical RMS of 0.15 mJy and an angular resolution of 5\arcsec \citep{Be95}. The FIRST survey area covers approximately 25\% of the total sky, out of which, approximately 80\% is in the north Galactic cap (8444 square degrees), and remaining 20\% is in the south Galactic cap (2131 square degrees) \citep{Be95}. The FIRST survey offers better resolution and sensitivity than the previous NRAO Very Large Array (VLA) Sky Survey (NVSS) at 1.4 GHz, which uses VLA D configuration and covers 82$\%$ of the celestial sphere with angular resolution 45\arcsec and an RMS of $\sim$ 0.45 mJy \citep{Co98}. The FIRST survey uses the NRAO VLA in its B configuration (VLA B). The FIRST sky is mapped with three minute snapshots covering a hexagonal grid using 2$\times$7 3 MHz frequency channels centered at 1365 and 1435 MHz until 2011\footnote{http://sundog.stsci.edu/first/description.html}. After 2011, the updated Expanded Very Large Array (EVLA) receiver came, which results in few changes in survey procedures. Frequencies of 1335 and 1730 MHz were used after 2011 with a one minute iteration using 2$\times$64 2 MHz frequency channels. The cleaning and calibration of the raw data are done using an automated pipeline based largely on routines in the Astronomical Image Processing System (AIPS).\footnote{http://info.cv.nrao.edu/aips/}. The latest version of FIRST data (2017 December 14) is used. This version covers R.A. = 07.0h to 17.5h, decl. = --08.0 deg to +57.6 deg in northern sky and R.A. = 20.4hr to  4.0hr, decl. = --11.5 deg to +15.4 deg in southern sky.

With the relatively high sensitivity and resolution of the FIRST survey, it is possible to study the morphology of faint radio galaxies in detail. The earlier FIRST database was used to search for radio galaxies with different distinct morphologies such as head tailed sources \citep{Mi19, Sa20}, hybrid morphology radio sources \citep{Go00}, compact steep spectrum sources and core-dominated triple sources (CSS; \citet{Ku02, Ma06}), giant radio sources (GRSs; \citet{Ku18}), and double-double radio galaxies (DDRG; \citet{Pr11}). Earlier, about 390 winged radio galaxy candidates were discovered using part of FIRST database \citet{Ch07, Ya19}. Here we are reporting a complete search of the FIRST database to look for radio galaxies with {\it X} and {\it Z} symmetries.

\subsection{Search Strategy}
\label{subsec:search-strategy}

We looked for {\it X}-shaped and {\it Z}-shaped radio sources using the VLA FIRST survey database. The catalog contains a total of 946,432 radio sources. We filtered all sources in the catalog that have an angular size of $>$10$^{\prime\prime}$ (i.e. at least twice the convolution beam size). Our filtering gives an output of 95,243 sources. We visually inspected fields of all the sources ($>$10$^{\prime\prime}$) to look for new candidate radio galaxies with wings. From the limited-resolution FIRST images ($\sim$ 5$^{\prime\prime}$), we confidently classified a sufficient number of objects as having the characteristic wing lengths of $>$80$\%$ of the active lobes. Following C07, we have also included galaxies with shorter wings ($< 80\%$ wing-to-lobe length ratios), as it is probable that projection effects are important. Since secondary jets are often diffused and weak and often lack bright peaks at the end (hotspots), it is difficult to measure the extent of these jets.

\begin{figure}
\includegraphics[width=6.2cm,angle=-90,origin=c]{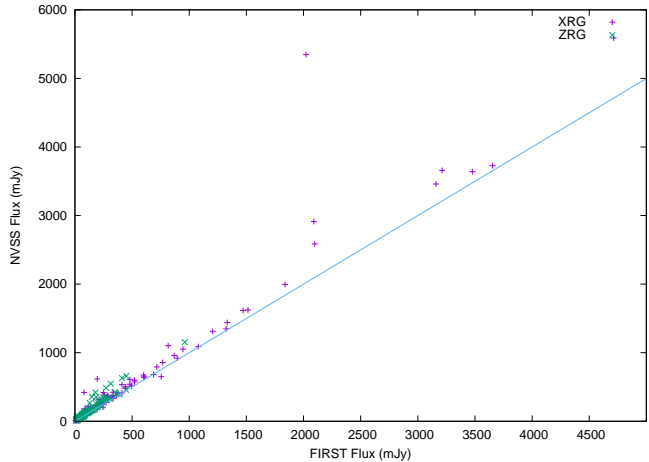}
    \caption{Flux densities measured from the FIRST catalog and the NVSS catalog are shown for all XRGs and ZRGs presented in this paper.}
    \label{fig:first-nvss}
\end{figure}

Earlier, C07 restricted search for {\it X}-shaped radio galaxies for fields with high dynamic ranges and $>15^{\prime\prime}$ source sizes. Y19 also restricted the search depending on the dynamic range. For the restricted search criterion, both of the studies missed many of the candidate {\it X}-shaped galaxies. C07 and Y19 manually searched 1648 and 5128 probable sources, in contrast to the probable 95,243 sources searched by the present paper.    

\subsubsection{Definition of XRGs and ZRGs}
\label{subsubsec:definition}

The wings in radio galaxies appear as the secondary lobes. Depending on the position of the wings, we classified the winged radio sources in two groups: XRGs and ZRGs. For the {\it "X"}-shaped radio sources, the wings (secondary lobes) appeared to be coming from the central spot or from near the central region. Here we define the near central region as the region covered by $\sim$ 25$\%$ of the primary jet length from the central spot. The sources are tagged as the {\it Z/S}-shaped radio source when the secondary lobes appear to come from the edges of the primary jet or the non-near central region. The morphology of sources depend on how many radio contours we consider and hence on the signal to noise. To make a uniform study, we started contours from 0.25 mJy for all images. 

\subsubsection{The Optical Counterparts and Properties}
\label{subsubsec:optical-counterpart}

We have searched for the optical/IR counterpart for each of the newly discovered XRGs and ZRGs from the Sloan Digital Sky Survey (SDSS) data catalog \citep{Gu06}, the Digital Sky Survey (DSS), and the NASA/IPAC Extra-galactic Database\footnote{https://ned.ipac.caltech.edu}. The optical/IR counterpart identification was based on the optical/IR source position relative to the radio galaxy morphology. We overlayed images from the FIRST survey with DSS2 red images. We have used the position of the optical/IR counterpart of XRGs and ZRGs as the position of these sources. Optical/IR counterparts are found for 98 sources out of a total of 161 sources for the case of {\it X}-shaped galaxies and 88 sources out of 135 sources for the case of {\it Z}-shaped sources, respectively. For objects with no clear optical/IR counterparts, we have used the location of the core of the radio galaxy or the intersection of both radio lobes as the position of these sources. We found that out of 135 redshift values, 123 are spectroscopic (91$\%$) and the rest 12 are photometric. Within 72 redshift measurements of XRGs, 67 are spectroscopic and 5 are photometric. Out of 63 redshift measurements of ZRGs, 56 are spectroscopic and 7 are photometric. Except otherwise stated, all redshifts are taken using SDSS catalog, release 12.

\section{Result}
\label{sec:result}
We report the discovery of 161 new XRGs and 135 new ZRGs. This discovery helps to significantly increase the number of this kind of sources. The larger samples of XRGs and ZRGs help us to do study various statistical properties of these sources, the result of which is described in the following subsections.  

All candidates of the XRGs and ZRGs are cataloged in Table \ref{Table:1} and \ref{Table:2}, respectively. For completeness, we have also included 21 XRG candidates and 44 ZRG candidates in these tables from \citet{Pr11}, which we independently identified following our method. In Figure \ref{fig:first-nvss}, the flux-density measurement of all XRGs and ZRGs are shown using FIRST and NVSS catalogs. For most of the sources, the flux density measured from the NVSS image is significantly higher than the corresponding measurement from the FIRST image. While the mean and median flux-density measurements from the NVSS catalog are 448.5 and 155 mJy, the mean and median flux densities from the measurements of the FIRST catalog are 381 and 122 mJy. Due to high resolution and lack of antennas in short spacing, the FIRST survey is prone to flux-density loss. For flux-density measurement of the sources reported in the present paper in 1400 MHz, we have used corresponding images from NVSS. The lower-resolution VLA configuration D was used for the NVSS (with a resolution of $\sim45\arcsec$), compared to B configuration of the FIRST survey (with a resolution of $\sim5\arcsec$), which means NVSS is better suited in detecting the most extended radio structure. NVSS counterparts are found for each of the sources, though due to less resolution, the {\it X}-shape and {\it Z}-shape are not evident from any of the NVSS images of these sources. For five XRGs (J0742+3339, J0932+1610, J1104+2828, J2218+0012 and J2324+1438) and two ZRGs (J0847+3147 and J1138+2039), we measure the 1400 MHz flux density from FIRST instead of NVSS as for these sources there were other background sources in the FIRST image inside the beam size of the NVSS image. 

In column 5 of Table 1 and Table 2, we have mentioned the catalog name from where the optical counterpart is found. The redshifts ($z$) of these sources are also mentioned when available (in column 6). We have also calculated the corresponding flux density of each source at 150 MHz (column 8) using TGSS \citet{In17}. We have calculated two-point spectral index between 150 and 1400 MHz $(\alpha_{150}^{1400})$ when flux density at 150 MHz is available and tabulated them in 9th column of Table 1 and 2. In columns 10 and 11, the linear size and luminosity of the sources are tabulated for the sources with known redshifts. In the last column, we have mentioned the name of other catalogs in radio wavelengths where these sources were mentioned earlier without detection of them as XRGs or ZRGs. 

\begin{figure*}
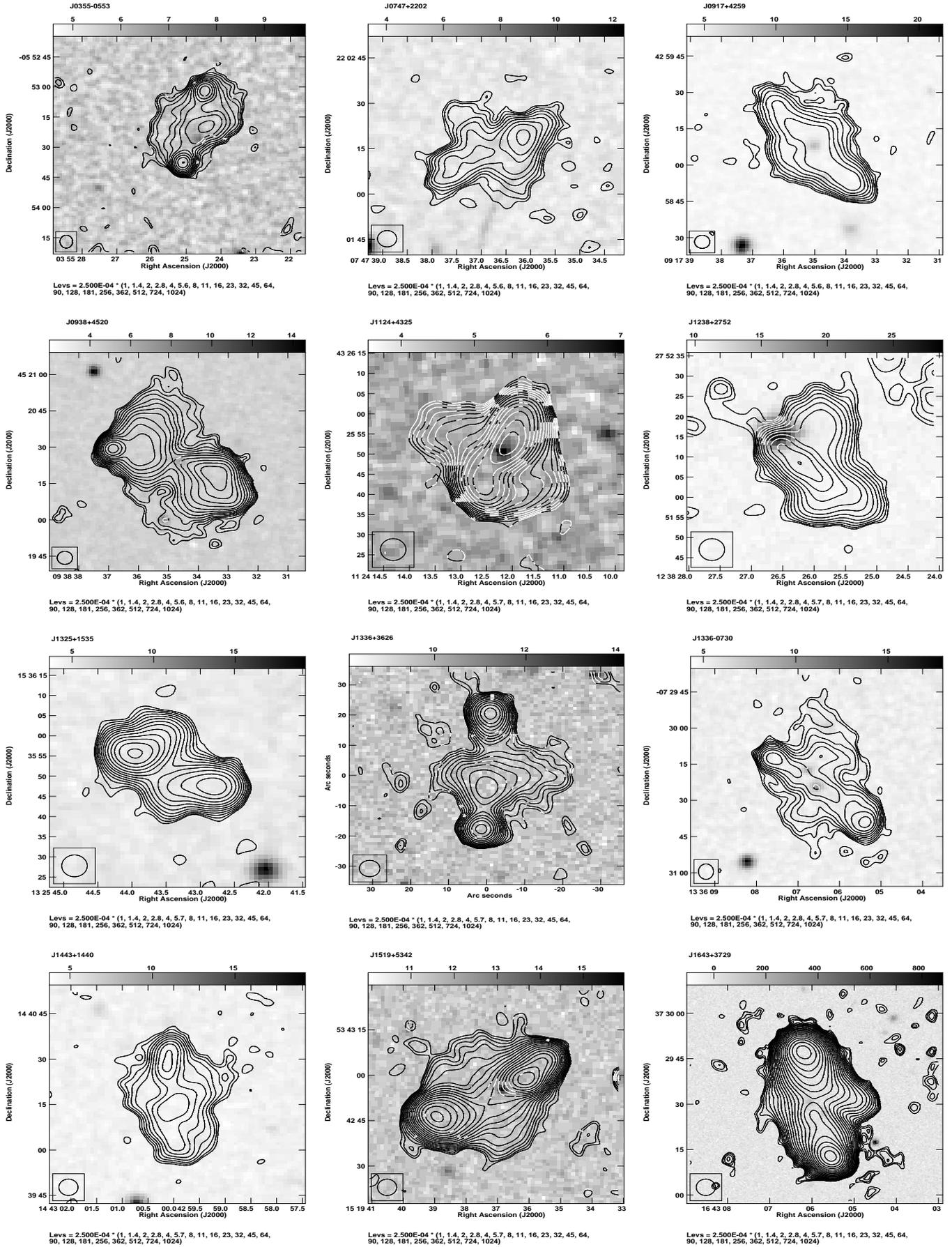

\vbox{
\centerline{
\includegraphics[height=6cm,width=6cm]{./J0355-0553.eps}
\includegraphics[height=6cm,width=6cm]{./J0747+2202.eps}
\includegraphics[height=6cm,width=6cm]{./J0917+4259.eps}}}
\vbox{
\centerline{
\includegraphics[height=6cm,width=6cm]{./J0938+4520.eps}
\includegraphics[height=6cm,width=6cm]{./J1124+4325.eps}
\includegraphics[height=6cm,width=6cm]{./J1238+2752.eps}}}
\vbox{
\centerline{
\includegraphics[height=6cm,width=6cm]{./J1325+1535.eps}
\includegraphics[height=6cm,width=6cm]{./J1336+3626.eps}
\includegraphics[height=6cm,width=6cm]{./J1336-0730.eps}}}
\vbox{
\centerline{
\includegraphics[height=6cm,width=6cm]{./J1443+1440.eps}
\includegraphics[height=6cm,width=6cm]{./J1519+5342.eps}
\includegraphics[height=6cm,width=6cm]{./J1643+3729S.eps}}}
\caption{The FIRST image of a sample of 12 {\it "X"}-shaped radio sources (contours) overlaid on the DSS2 red image (grey scale).}
	\label{XRG}
\end{figure*}

In Figure \ref{XRG}, we have shown the example of 12 XRGs, and in Figure \ref{ZRG}, we have shown the example of 12 ZRGs. The DSS2 red optical images are overlayed with the FIRST radio images. Synthesized beams are shown in one corner of the images. 

%----------------------------------------------------------
\begin{figure*}
\vbox{
\centerline{
\includegraphics[height=6cm,width=6cm,origin=c]{./J0219+0155.eps}
\includegraphics[height=6cm,width=6cm,origin=c]{./J0814+5059.eps}
\includegraphics[height=6cm,width=6cm,origin=c]{./J0816+0458.eps}}}
\vbox{
\centerline{
\includegraphics[height=6cm,width=6cm,origin=c]{./J0831+0125.eps}
\includegraphics[height=6cm,width=6cm,origin=c]{./J0902+5203.eps}
\includegraphics[height=6cm,width=6cm,origin=c]{./J1215+1709.eps}}}
\vbox{
\centerline{
\includegraphics[height=6cm,width=6cm,origin=c]{./J1308-0500.eps}
\includegraphics[height=6cm,width=6cm,origin=c]{./J1410+3014.eps}
\includegraphics[height=6cm,width=6cm,origin=c]{./J1418+3530.eps}}}
\vbox{
\centerline{
\includegraphics[height=6cm,width=6cm,origin=c]{./J1442+5043.eps}
\includegraphics[height=6cm,width=6cm,origin=c]{./J1540+1110.eps}
\includegraphics[height=6cm,width=6cm,origin=c]{./J1643+2642.eps}}}
\caption{The FIRST image of a sample of 12 {\it "Z"}-shaped radio sources (contours) overlaid on the DSS2 red image (grey scale).}
	\label{ZRG}
\end{figure*}

\subsection{Spectral Index ($\alpha$)}
\label{subsec:spectral-index}

\begin{figure*}
\includegraphics[width=6cm,angle=-90,origin=c]{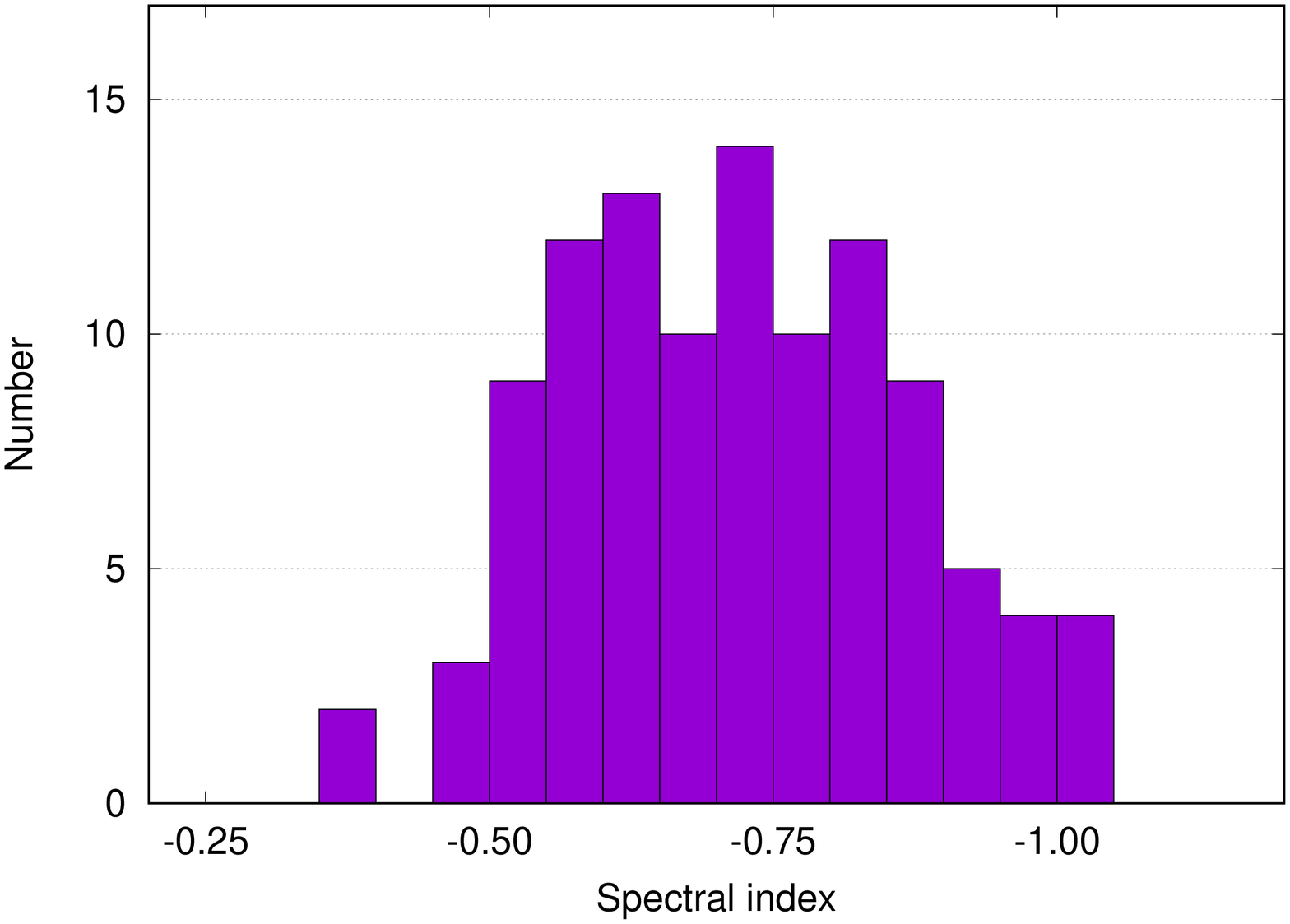}
\includegraphics[width=6cm,angle=-90,origin=c]{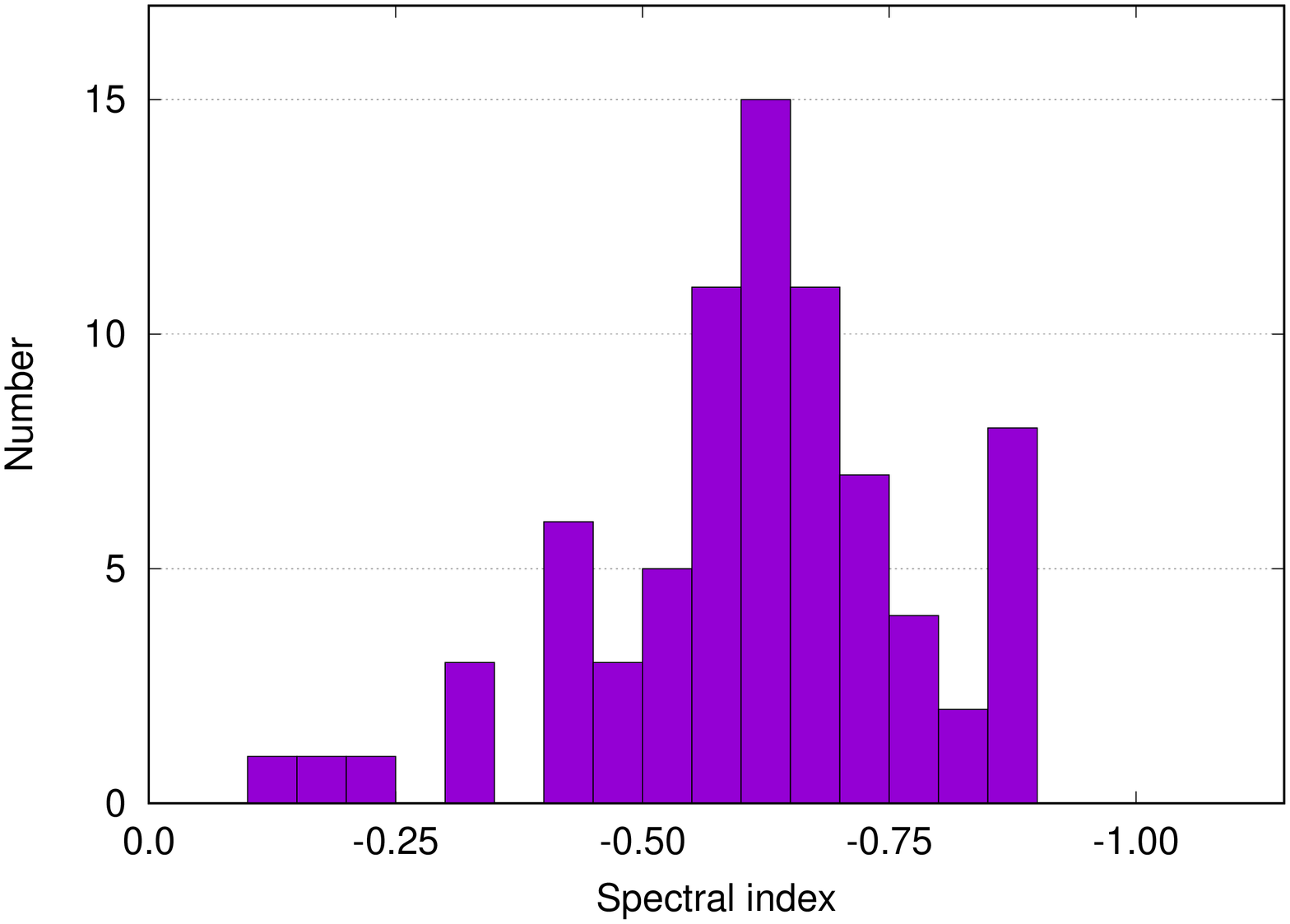}
	\caption{Histogram showing the spectral index distribution of radio galaxies presented in the current paper with {\it "X"} (left) and {\it "Z"} shapes (right).}
    \label{fig:hist-alpha}
\end{figure*}

We have calculated the two-point spectral index of newly discovered sources between 150 and 1400 MHz (assuming $S_{\nu} \propto \nu^\alpha$, where $S_{\nu}$ is the radiative flux density at a given frequency, $\nu$, and $\alpha$ is the spectral index) and mentioned them in Table 1 and Table 2. Spectral index measurements are available for 107 XRGs and 79 ZRGs. For the rest of the sources, due to higher RMS in the images of TGSS, the sources were not detectable in 150 MHz map of TGSS.

Figure \ref{fig:hist-alpha} shows histogram with the spectral index distribution of sources presented in the current article for XRGs (left) and ZRGs (right). The plot shows that the total span of $\alpha_{150}^{1400}$ is from $-0.35$ to $-1.05$ for XRGs and $-0.14$ to $-1.21$ for ZRGs. Among XRGs, J2249+0209 has the lowest spectral index with $\alpha^{1400}_{150}=-1.05$ and J0758+4406 has the highest spectral index with $\alpha^{1400}_{150}=-0.35$. Among ZRGs, J1325+5736 have the lowest spectral index with $\alpha^{1400}_{150}=-1.21$ and J1524+1627 has the highest spectral index with $\alpha^{1400}_{150}=-0.14$. Most of the radio galaxies show steep radio spectrum $\alpha\le-0.5$, as expected from the lobe dominated radio source. There is a clear distinction between peaks in the histogram for XRGs and ZRGs. For XRGs, the histogram shows two peaks near $-0.55$ and $-0.7$, and for ZRGs, the histogram shows a peak near --0.65. 
\subsection{Radio Luminosity ($L_{rad}$)}
\label{subsec:lum}

The radio luminosities ($L_{rad}$) of newly discovered XRGs and ZRGs are calculated using

\begin{eqnarray}
L_{rad} = \nonumber 1.2\times10^{27}D^2_{\text{Mpc}}S_0{\nu_0^{-\alpha}}(1+z)^{-(1+\alpha)}          \\
\nonumber \times(\nu_u^{(1+\alpha)}-\nu_l^{(1+\alpha)})(1+\alpha)^{-1} \text{erg s$^{-1}$} \\
\label{eqn:1}
\end{eqnarray}
where $D_{\text{Mpc}}$; is luminosity distance to the source (Mpc), $S_0$ is the flux density (Jy) at a given frequency, $\nu_0$ (Hz); $z$ is the redshift of the radio galaxy; and $\alpha$ is the spectral index ($S \propto \nu^\alpha$). $\nu_u$ (Hz) and $\nu_l$ (Hz) are the upper and lower cutoff frequencies \citep{Od87}.
In our calculation, we assume the upper and lower cutoff frequencies as 100 GHz and 10 MHz, respectively.

\begin{figure}
\includegraphics[width=6cm, angle=-90,origin=c]{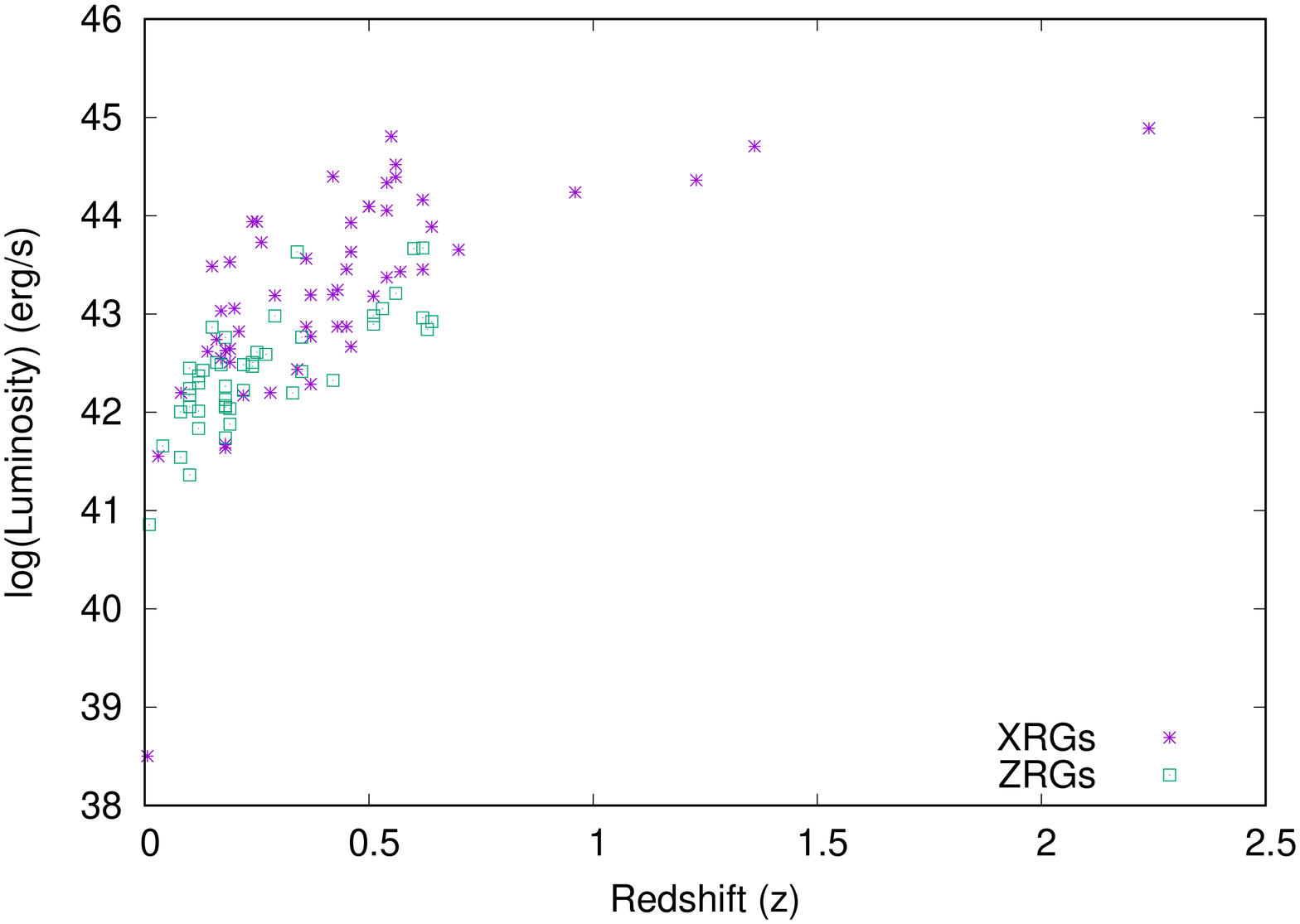}
	\caption{Distribution of radio luminosity ($L_{rad}$) with red shift ($z$) for XRGs and ZRGs.}
    \label{fig:lum-redshift}
\end{figure}

In Figure \ref{fig:lum-redshift}, we plot the distribution of radio luminosities of XRGs and ZRGs presented in the current paper with known redshifts ($z$). We have performed a Kolmogorov-Smirnov test \citep{Pe83, Sm48} between the distribution of XRGs and ZRGs and found they are consistent with each other with $D=0.52$ and $p$-value$=8\times10^{-7}$. The radio luminosity of the sources at 1.4 GHz is in the order of $10^{43}$ erg s$^{-1}$, which is bit less than a typical radio galaxy. The average value of Luminosity in \citet{Ni93}, which is a collection of 540 double radio sources, is $Log~L$ [erg s$^{-1}$] is 44.07. The average value of $Log~L$ is higher for XRGs compared to ZRGs. For XRGs, the average value of $Log~L$ [erg s$^{-1}$] is 43.24 (1$\sigma$ standard deviation=1.05, median=43.31) and that of ZRGs is 42.42 (1$\sigma$ standard deviation=0.58, median=42.42).  

The most luminous XRG in our sample is J1124+4325 with $L_{rad}=7.76\times10^{44}$ erg s$^{-1}$ and the most luminous ZRG in our sample is J1156+2138 with $L_{rad}=4.70\times10^{43}$ erg s$^{-1}$. The least luminous XRG in our sample is J1354+5840 with $L_{rad}=3.19\times10^{38}$ erg s$^{-1}$ and the least luminous ZRG in our sample is J1140+1743 with $L_{rad}=7.2\times10^{40}$ erg s$^{-1}$. 

\subsection{Properties of XRGs}
\label{subsec:prop-XRG}

\begin{figure}
\includegraphics[width=6cm,angle=-90,origin=c]{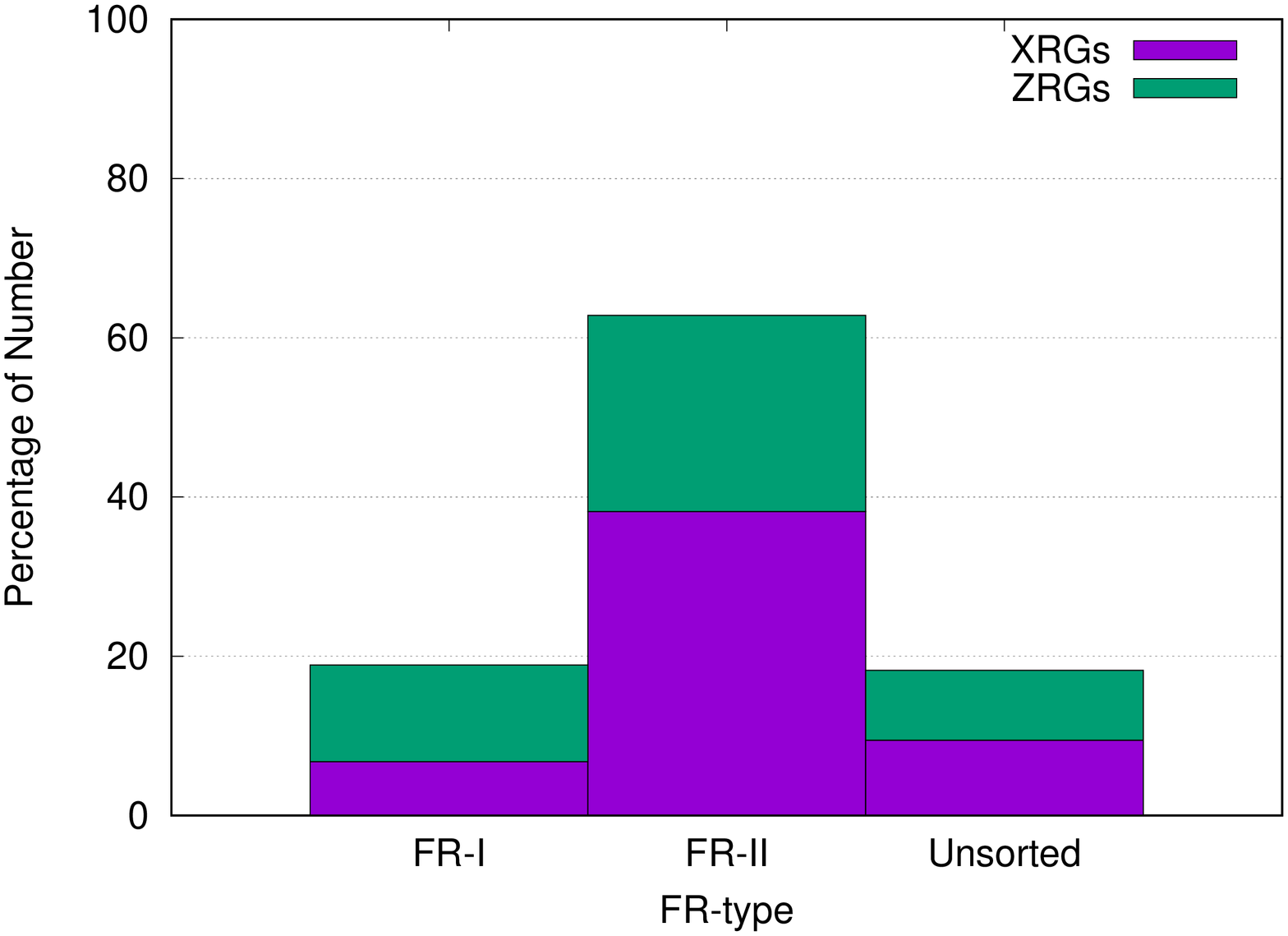}
	\caption{Histogram showing the number of FR I and FR II sources amongst the XRGs and ZRGs reported in the present paper.}
    \label{fig:FRI-FRII}
\end{figure}

Most of the XRGs are found to be FR II radio galaxies \citep{Le92, Ch07, Ya19}. This trend is strengthened by the present catalog of XRGs. Out of 161 XRGs presented in the current paper, 70\% (113) are FR II radio galaxies and 13\% (20) are FR I radio galaxies. For 28 XRGs, the jet morphology is complex and could not be classified. For the uniform classification of sources, we used radio maps starting from 0.25 mJy contour. Depending on the position of peak flux in lobes of the major axis, we classified the source as FR I and FR II. Figure \ref{fig:FRI-FRII} shows a histogram indicating the number of XRGs with FR I and FR II type. For none of the XRGs in the present sample, the secondary lobes are FR II type.

\subsubsection{Major and Minor Axes}
\label{subsubsec:maj-axis}
\begin{figure}
\includegraphics[width=6cm,angle=-90,origin=c]{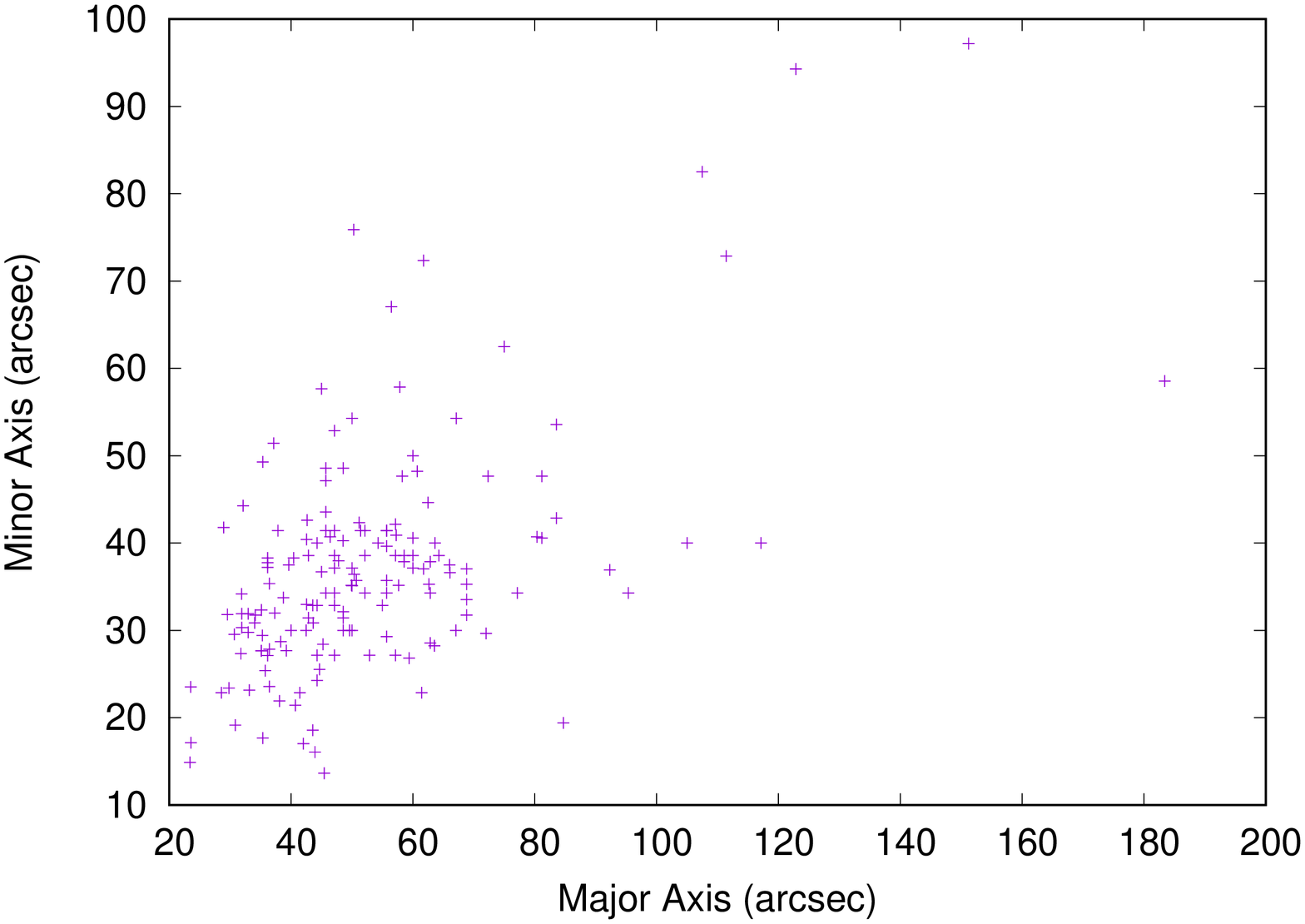}
    \caption{The distribution of the size of major and minor axes of the discovered radio galaxies with an {\it X}-like shape.}
    \label{fig:major-minor-lobe}
\end{figure}

In Figure \ref{fig:major-minor-lobe}, the angular size of the major axis of X-shaped radio galaxies for the angular size of the corresponding minor axis is plotted. The angular sizes depend on the lowest contour of the image. For uniformity, we have started contours from 0.25 mJy for images of all XRGs. For 27 galaxies in our list, the minor axis is elongated in only one direction. These 27 galaxies are not included in Figure \ref{fig:major-minor-lobe}.

\begin{figure}
\includegraphics[width=6cm,angle=-90,origin=c]{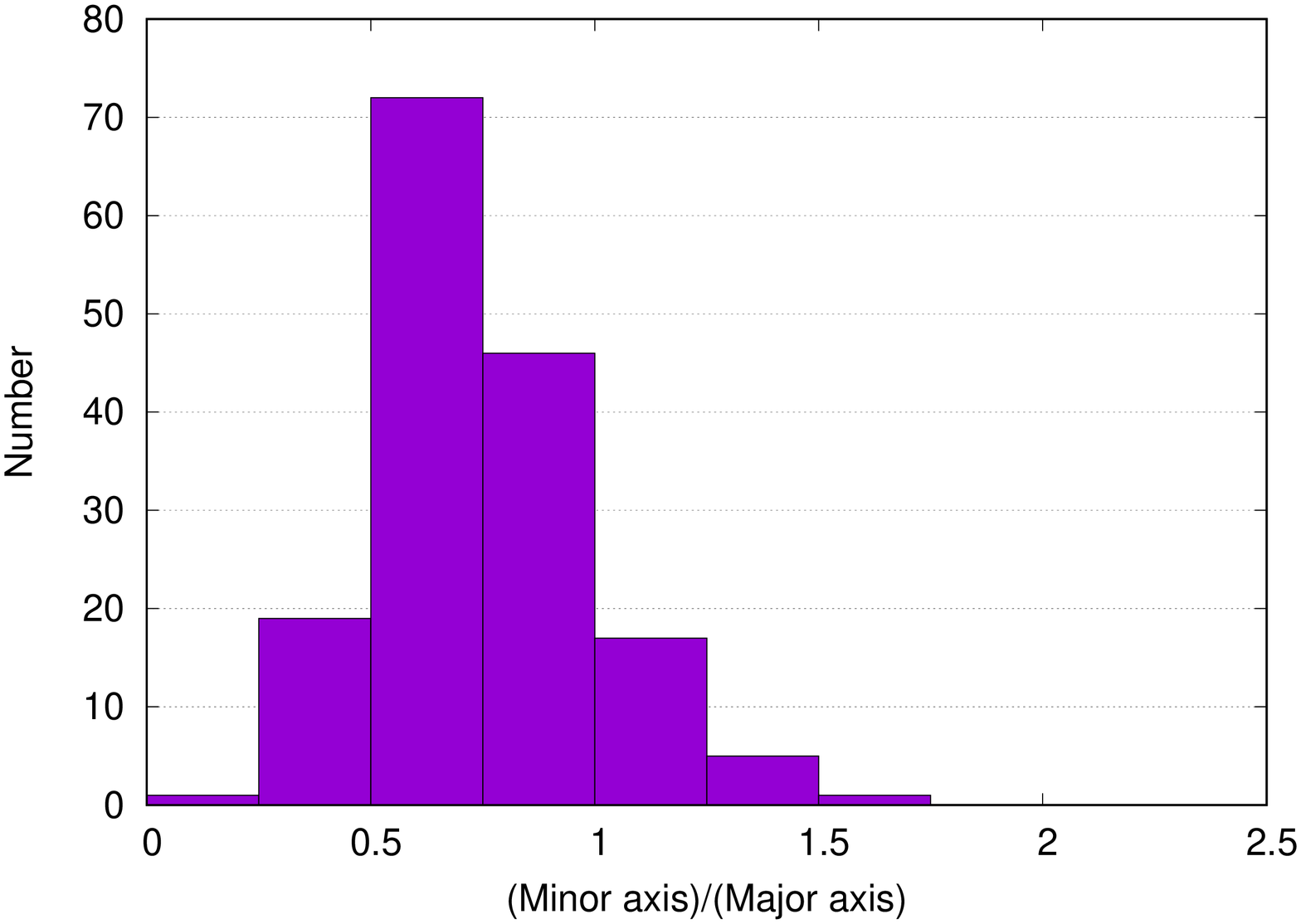}
    \caption{Histogram showing the distribution of size between major and minor axis in `X' shaped sources. }
    \label{fig:jetratio}
\end{figure}

In Figure \ref{fig:jetratio}, we have shown a histogram with the distribution of ratio between the minor and major axes of radio galaxies with {\it "X}-like shape", as presented in this paper. Out of 161 radio galaxies included in the Figure, 47 radio galaxies has secondary jets with $50-75\%$ of the size of the primary jets. For 32 radio galaxies, the secondary jets are with $75-100\%$ size of the primary jet. For seven XRGs, the secondary jets are bigger than the primary jets. Here we should remember that primary jets also could be affected by the effect of projection.

\subsubsection{Angle between the Major and Minor Axes}
\label{subsubsec:axis-angle}

\begin{figure}
\includegraphics[width=6cm,angle=-90,origin=c]{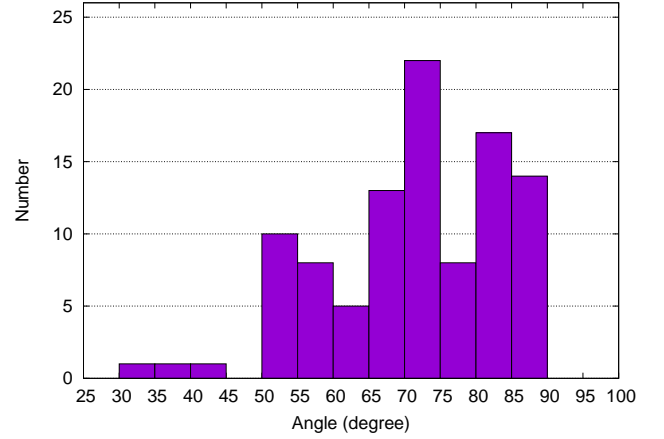}
    \caption{Histogram showing the distribution of angle between major and minor axes in {\it "X"}-shaped sources. }
    \label{fig:X-angle}
\end{figure}

Figure \ref{fig:X-angle} shows a histogram with the distribution of the angle between major and minor axes in XRGs. The angle between the intersection of two virtual lines with peak fluxes in each lobe of major and minor axes is used to measure the angle between two axes. Sources with one-sided wings are not included in the figure. Sources with diffused and nonlinear secondary lobes are also not included in the study to avoid error in angle measurement. Total 161 XRGs are used in the figure. Majority of the sources (63\%) have the angle in the range of $(70 - 90)$ degrees. The histogram shows a peak in the number of radio galaxies at 75 degrees.

\subsection{Properties of the ZRGs}
\label{subsec:prop-ZRG}

Out of 135 ZRGs presented in the current paper, 54\% (36) are FR II radio galaxies and 27\% (73) are FR I radio galaxies. For 26 ZRGs, the jet morphology is complex and could not be classified.

\subsubsection{The Angular Size}
\label{subsubsec:angular-size}

\begin{figure}
\includegraphics[width=6cm,angle=-90,origin=c]{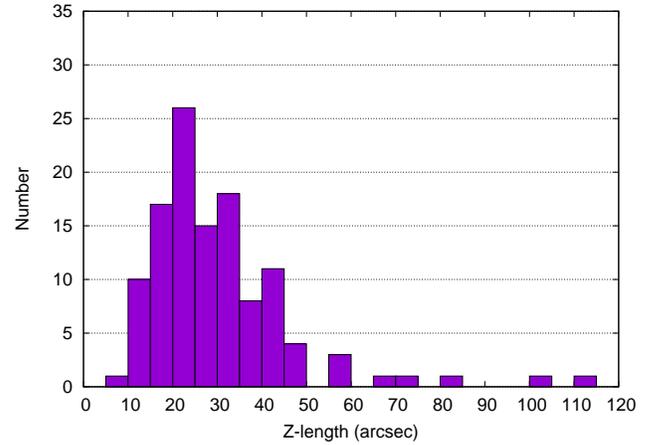}
    \caption{Histogram showing the distribution of the angular size of {\it "Z"}-shaped sources. }
    \label{fig:Z-length}
\end{figure}

In Figure \ref{fig:Z-length}, the distribution of the angular size of primary jets in ZRGs is shown. Here we have measured the angular size of the primary jet in between the two oppositely directed secondary wings. The angular sizes depend on the lowest contour of the image. For uniformity, we have started contours from 0.25 mJy. The angular size is determined by the task `tvdist' in AIPS.

For most of the sources, the angular size of primary jets is less than 1 arcmin. The histogram shows a sharp peak near 20 arcsec. There are only two sources with a primary jet angular size of more than 90 arcsec. J1417+0812 has a primary jet angular size of 113 arcsec and J1303+0339 has an angular size of 104 arcsec. The ZRG with the smallest primary jet angular size is J2137--0811 (9 arcsec).

Amongst the sources with known redshift, the source with the highest primary jet linear size is J1319+0502 (1.41 Mpc). J1303+0339 and J1536+2357 have a primary jet linear sizes of 0.47 and 0.44 Mpc, respectively. J1140+1743 is the smallest source with the primary jet linear size of 5.85 kpc. The primary jet linear size of J0219+0155 is 37.8 kpc.

\section{Discussion}
\label{sec:disc}

Out of a total of 161 XRGs and 135 ZRGs, redshifts are known for 72 and 63, respectively. Amongst the XRGs, $0<z<0.5$ for 53 sources and $z > 0.5$ for 19 sources. J1124+4325 showed the highest redshift with $z = 2.245$ \citep{Ri09}. Spectroscopically, J1124+4325 is classified as a quasar \citep{Xu14}. J1333+0219 and J0750+1144 showed redshifts of $z = 1.228$ \citep{Ni98} and $z=0.955$ \citep{Ri09}, respectively. Amongst the ZRGs, $0 < z < 0.5$ for 53 sources and $z > 0.5$ for 10 sources. Amongst the ZRGs, J1319+0502 has the highest redshift at $z = 1.285$ \citep{Ri09}. J1124+4325 and J1319+0502 are the farthest known XRG and ZRG, respectively. Previously, J0229+132 ($z=2.065$) was the farthest {\it X}-shaped radio source candidate with its large core dominance \citep{Ma06}. The highest redshift XRG in \citet{Ch07} is J1206+3812 ($z=0.838$). From the optical spectroscopy of {\it X}-shaped and {\it Z}-shaped sources reported earlier, it is found that a good fraction of the sources are quasars \citep{Ch07}. Since number density of quasars peak at $z\sim1-2$ \citep{Bo00}, we expect a good number of XRGs and ZRGs in these redshifts but since minor jets in faint sources (which are often diffused) are not detectable for many of the sources, we are missing many of the XRGs and ZRGs with $z>1$. 

For the XRG J0057--0123, two supernovae are found inside the radio map. SN 2010jo and SN 2007nq \citep{Ha12} are located at 11.7 and 21.6 arcsec far from the center of the galaxy at one of the major lobes of the galaxy.

All the XRGs, with flux densities $>2$ Jy, are FR II radio galaxies. Amongst XRGs, J1001+2847 is the brightest radio galaxy with flux density $F_{1400}$ = 5589 mJy. J1001+2847 is a Seyfert galaxy \citep{Ve06}. J0057--0123 ($F_{1400}$ = 5384 mJy), J0655+5408 ($F_{1400}$ = 3727 mJy), J1422+1935 ($F_{1400}$ =3659 mJy), J0950+1420 ($F_{1400}$ = 3639 mJy), J1617+3222 ($F_{1400}$ = 3222 mJy), and J1235+2120 ($F_{1400}$ = 2013 mJy) have flux densities more than 2 Jy. Only one ZRG has a flux density greater than 1 Jy. J1140+1743 is the brightest ZRG with flux density 1154 mJy.

Though the present paper, along with C07 and Y19, report the discovery of a large number of XRGs and ZRGs from the FIRST survey, it should be remembered that, due to high resolution in the FIRST image and the absence of antennas with shorter baseline, a large number of XRGs and ZRGs with larger sizes and diffused emission will be missed. 

As mentioned earlier, the wings in XRGs and ZRGs may be significantly shortened due to projection effects. The flux-density asymmetry between two sides of the wings, as found in some XRGs, may indicate that the wings are close to our line of sight and flux-density asymmetry is due to Doppler boosting. Some of the 26 sources with one-sided wing may be an extreme example of Doppler boosting.  

We have identified 161 XRGs and 135 ZRGs, making the ratio of XRGs to ZRGs equal to 1.2. Earlier, \citet{Sa18} pointed out using a sample of \citet{Ch07} that the ratio of XRG to ZRG in their sample is $\sim2$. The current paper detected more ZRGs compared to \citet{Ch07} due to the fact that ZRGs are statistically fainter than XRGs (see Section \ref{subsec:spectral-index}) and \citet{Ch07} missed many of the ZRGs because of its dynamic range specific selection criterion. \citet{Sa18} also pointed out that their sample has only one FR I sources (2.7\%) out of 37 XRGs. Again, the present paper detected more FR I sources than that reported in \citet{Sa18} because FR I sources are statistically fainter than FR II sources and the present paper did not make any restrictions with respect to the dynamic range filtering out relatively fainter sources. The present paper shows that though most of XRGs are still FR II in the present sample (70\%) FR I XRG sources are also not negligible (13\%) contrary to the findings of \citet{Sa18}.

A total of 127 sources, i.e. 43\% sources in our sample are inversion symmetric. Out of the 127 sources, 58 XRGs and 69 ZRGs are symmetric. The inversion symmetric structure hints at a host-related cause for the formation of wings.

To explain the nature of XRGs, the backflow of plasma is one of the widely accepted models. The buoyancy-driven backflow cannot propagate faster than the external sound speed \citep{Le92} and so the wings produced by the buoyancy forces are not expected to be longer than the main radio lobes, which are known to advance supersonically. We have found seven sources where wings are longer than the primary axes. \citet{Sa18} also found two such sources within their sample of 87 sources. So, for some of the XRGs, backflow may still occur but it is not the only reason for all of the XRGs (especially for sources with longer wings). But it is still possible that the alignment effect plays a roll in the primary jets for which it looks shorter than the secondary jets. 

Multifrequency high-resolution observation of these sources is encouraged to confirm the nature of these sources. Discovery of a large number of XRGs and ZRGs shows that these kinds of sources are not rare, and with a future deeper high-resolution survey, we expect to discover more such objects.

\section{Conclusion}
\label{sec:conclusion}

We have discovered a total of 296 radio sources with wings, out of which 161 are XRGs and 135 are ZRGs. This discovery helps to increase the number of known XRGs and ZRGs significantly. Optical/IR counterparts of most of the XRGs and ZRGs are identified. As expected, most of the XRGs and ZRGs show a steep spectral index between 150 MHz and 1400 MHz. Most of the XRGs are FR II radio galaxies. The average value of luminosities for XRGs is found to be higher than that of ZRGs. 

We thank the anonymous reviewer whose suggestion help to improve the quality of the paper. This research has made use of the NASA/IPAC Extra-galactic Database (NED), which is operated by the Jet Propulsion Laboratory, California Institute of Technology, under contract with the National Aeronautics and Space Administration. T.K.S. and S.M. want to acknowledge funding from Rashtriya Uchchatar Shiksha Abhijan 2.0. 

\bibliographystyle{aasjournal}

\begin{table*}
\caption{\bf Candidate {\it X}-Shaped Radio Sources}
\begin{footnotesize}
\begin{centering}
\begin{tabular}{cccccccccccl}
\hline
Catalog &  Name         & R.A.          & Decl.         & Ref.     &Redshift&$F_{1400}$&$F_{150}$&$\alpha_{150}^{1400}$&Linear&$L$&Other Catalogs \\
Number  &               & (J2000.0)     & (J2000.0)     &    & ($z$)    & (mJy)    & (mJy)        &                    &Size& (ergs$^{-1}$)  &    \\
	&               &               &               &       &        &          &         &     			&(kpc)& ($\times 10^{42}$) &  \\
% (1)   &   (2)         & (3)           &  (4)          & (5)    &  (6)   & (7)      & (8)        & (9)        & (10)        & (11)          & (12)    \\
\hline            
~~1     & J0017--0149   & 00 17 32.20   & --01 49 25.9 	&GALEXMSC& ---  & ~~91	& ---	& ---	&---& ---	& 1, 2, 8 \\
~~2	& J0022--0807	& 00 22 55.64 	& --08 07 52.4	& SDSS	& 0.44  & ~~98	& ---	& ---	&~796& ---	& 1 \\
~~3	& J0023--0915	& 00 23 25.26	& --09 15 38.2	& SDSS	& ---   & ~117	& ---	& ---	&---& ---	& 1 \\
~~4	& J0057--0123	& 00 57 34.92	& --01 23 27.9	& SDSS	& 0.04  & 5348	& ---	& ---	&~137& ---	& 2, 3, 8, 9, 10, 11, 14, 24, 29 \\
~~5	& J0110--0924	& 01 10 34.36	& --09 24 45.2	& EE	& ---   & ~135	& ---	& ---	&---& ---	& 1, 2, 8, 10 \\
~~6	& J0112--0804	& 01 12 42.41	& --08 04 02.4	& EE	& ---   & ~365	& ---	& ---	&---& ---	& 1, 2, 8, 10 \\
	~~7	& J0139+0131	& 01 39 57.25	& +01 31 46.2	& GALEXASC& 0.26$^{30}$& 1312	& ~4750	& --0.58&~415& ~53.65& 1, 2, 4, 8, 9, 10, 11, 12 \\     
~~8	& J0144+1212	& 01 44 36.38	& +12 12 54.8	& EE	& ---   & ~~11	& ~~~38	& --0.56&---& ---	& 1	\\
~~9	& J0145--0941	& 01 45 03.87	& --09 41 16.6	& SSTSL2& ---   & ~~90	& ---	& ---	&---& ---	& ---	\\
~10	& J0152--0759	& 01 52 39.32	& --07 59 13.8	& SDSS	& 0.41  & ~~79	& ---	& ---	&~772& ---	& ---	\\
~~~11$^a$& J0212--0450	& 02 12 16.55	& --04 50 38.8	& EE	& ---   & ~157	& ---	& ---	&---& ---	& 8	\\
~12	& J0335--0719	& 03 35 48.23	& --07 19 12.5	& EE	& ---   & ~~86	& ---	& ---	&---& ---	& 1, 2 \\
~13	& J0354--0522	& 03 54 04.69	& --05 22 08.9	& EE	& ---   & ~608	& ---	& ---	&---& ---	& 1, 2, 4, 8, 9, 10 \\
~14	& J0355--0553	& 03 55 24.70	& --05 53 19.8	& EE	& ---   & ~~80	& ---	& ---	&---& ---	& 1   \\
~15	& J0654+5814	& 06 54 22.91	& +58 14 26.0	& EE	& ---   & ~~95	& ~~621	& --0.84&---& ---	& ---	\\
~16	& J0655+5408	& 06 55 14.73	& +54 08 57.2	&GALEXASC& 0.24$^{31}$ & 3727	& 31666	& --0.96&~320& ~86.98	& 1, 2, 3, 10, 12 \\
~17	& J0703+6014	& 07 03 11.18	& +60 14 23.1	& 2MASX	& ---   & ~~55	& ~~154	& --0.46&---& ---	& 1	\\
~18	& J0709+5716	& 07 09 58.26	& +57 16 51.6	& EE	& ---   & ~271	& ~1660	& --0.81&---& ---	& 1, 2, 6, 10, 12 \\
~19	& J0712+5430	& 07 12 45.08	& +54 30 10.8	& EE	& ---   & ~381	& ~2406	& --0.82&---& ---	& 1, 2, 6, 10, 12 \\
~20	& J0719+5519	& 07 19 36.64	& +55 19 42.8	& EE	& ---   & ~~61	& ~~424	& --0.87&---& ---	& 1		\\
~21	& J0721+3551	& 07 21 29.03   & +35 51 38.2	& SDSS	& ---   & ~503	& ~4799	& --1.01&---& ---	& ---		\\
~22	& J0723+3323	& 07 23 59.55	& +33 23 20.2	& EE	& ---   & ~148	& ~~525	& --0.57&---& ---	& 1, 2, 6, 12, 20 \\
~~~23$^a$& J0742+3339	& 07 42 59.47	& +33 39 52.7	& SDSS	& 0.64  & ~334$^\ast$& ~2729& --0.94&1214& ~77.05	& ---	\\
~24	& J0743+1733	& 07 43 24.28	& +17 33 41.3	& EE	& ---   & ~423	& ~2494	& --0.79&---& ---	& 1, 2, 10, 12 \\ 
~25	& J0747+2202	& 07 47 36.74	& +22 02 15.9	& SDSS	& 0.46  & ~~44	& ~~301	& --0.86&~606& ~~4.65	& 1	 \\
~26	& J0748+2324	& 07 48 45.10	& +23 24 45.8	& SDSS	& 0.19  & ~515	& ~3238	& --0.82&~260& ~33.87	& ---	\\
~27	& J0749+2007	& 07 49 19.08	& +20 07 53.7	& SDSS	& 0.37  & ~139	& ~~960	& --0.86&~518& ~~1.94	& 1, 2, 10, 12 \\
~28	& J0750+1144	& 07 50 25.95	& +11 44 52.0	& SDSS	& 0.96  & ~289	& ~1846	& --0.83&1512& 172.51& 15	\\
~29	& J0758+4406	& 07 58 08.43	& +44 06 17.0	& EE	& ---   & ~~31	& ~~~68	& --0.35&---& ---	& 1		\\
~30	& J0758+1946	& 07 58	19.52	& +19 46 56.5	& SDSS	& ---  	& ~~51	& ~~421	& --0.94&---& ---	& 1		\\
~31	& J0758+1020	& 07 58 26.24	& +10 20 18.6	& SDSS	& 0.37 	& ~213	& ~~977	& --0.68&~447& ~15.63 & 1, 2, 12 \\
~32	& J0802+6353	& 08 02 02.34	& +63 53 03.3	& LQAC	& 0.47 	& 1052	& ---	& ---	&1386& ---	& 2, 6, 10, 12 \\
~~~33$^a$& J0804+4659	& 08 04 53.99	& +46 59 57.0	& EE	& ---  	& ~~52	& ---	& ---	&---& ---	& 1		\\
~34	& J0811+2954	& 08 11 11.85	& +29 54 53.6	& EE	& ---  	& ~162	& ~1550	& --1.01&---& ---	& 1, 2, 12, 20 \\
~35	& J0817+2951	& 08 17 03.62	& +29 51 47.1	&TONS08w& 0.37$^{32}$ 	& ~~93	& ~~686	& --0.89&~316& ~~5.88	& 1, 10 \\
~36	& J0823+0335	& 08 23 16.90	& +03 35 36.5	& EE	& ---  	& ~~90	& ~~214	& --0.39&---& ---	& 1		 \\
~37	& J0825+4623	& 08 25 24.05	& +46 23 33.8	& EE	& ---  	& ~~38	& ~~148	& --0.61&---& ---	& ---		\\
~38	& J0827+3748	& 08 27 05.80	& +37 48 42.2	& EE	& 0.21 	& ~379	& ~3644	& --1.01&~446& ~~6.65& 1, 2, 4, 10, 12, 20 \\
~39	& J0828+3057	& 08 28 03.00	& +30 57 42.8	& SDSS	& 0.36 	& ~102	& ~~460	& --0.67&~440& ~~7.39& 1, 2, 12, 20 \\
~~~40$^a$& J0830+3510	& 08 30 36.26	& +35 10 59.2	& SDSS	& ---  	& ~~76	& ---	& ---	&---& ---	& 1, 2	\\
~41	& J0840+6413	& 08 40 25.47 	& +64 13 17.7	& EE	& ---  	& ~~30	& ~~213	& --0.88&---& ---	& 1		\\
~42	& J0849+0949	& 08 49 40.02	& +09 49 21.2	& SDSS	& 0.36 	& ~575	& ~3532	& --0.81&~816& ~36.52& 2, 4, 8, 9, 10, 11, 15 \\
~~~43$^a$& J0857--0339	& 08 57 11.73	& --03 39 41.0	& EE	& 0.17 	& 1103	& ---	& ---	&~405& ---	& 1, 2, 4, 8, 9, 10, 11, 15 \\
~44	& J0858+5740    & 08 58 03.85   & +57 40 13.8 	& SDSS  & 0.45 	& ~~61	& ~~253	& --0.64&~427& ~~7.45& 1, 6, 12 \\
~45 & J0906+1646 & 09 06 32.53 & +16 46 05.6 & SDSS & 0.41  & 1616 & --- & --- &~818& --- & 3 \\
~46	& J0914+6121	& 09 14 00.60	& +61 21 11.6	& EE	& ---  	& ~~22	& ~~110	& --0.72&---& ---	& ---		\\
~47	& J0917+4259	& 09 17 35.04	& +42 59 07.6	& SDSS	& ---  	& ~~81	& ~~346	& --0.65&---& ---	& 1		\\
~48	& J0931--0540	& 09 31 41.67	& --05 40 48.0	& EE	& ---  	& ~~10	& ---	& ---	&---& ---	& 1	\\
~49	& J0932+1610	& 09 32 43.96	& +16 10 53.1	& EE	& --- 	& ~273$^\ast$& ---& ---	&---& ---	& 1, 2 \\
~50	& J0933--0507	& 09 33 37.48	& --05 07 18.8	& EE	& ---  	& ~~90	& ---	& ---	&---& ---	& ---	 \\
~51	& J0938+4520	& 09 38 34.68	& +45 20 23.7	& SDSS	& 0.45 	& ~258	& ~1286	& --0.72&~906& ~28.51& 1, 2, 6, 10, 12, 21 \\
~52	& J0941+3853	& 09 41 04.00	& +38 53 50.9	& SDSS	& 0.62 	& ~682	& ~4176	& --0.81&1325& 144.90& 1, 2, 4, 10, 20 \\
~53	& J0950+1420	& 09 50 10.79	& +14 20 00.6	& SDSS	& 0.55	& 3639	& 18298	& --0.72&~984& 640.99& 1, 2, 3, 11, 15 \\
~~~54$^a$& J1001+2847	& 10 01 49.52 	& +28 47 08.9	& SDSS	& 0.18 	& 5589	& ---	& ---	&~561& ---	& 2, 4, 10, 11, 20 \\
~55	& J1021+4425	& 10 21 16.98	& +44 25 39.8	& EE	& ---  	& ~~94	& ---	& ---	&---& ---	& ---	\\
~56	& J1022+5213	& 10 22 12.66	& +52 13 42.4	& EE	& ---  	& ~151	& ~1385	& --0.99&---& ---	& 1, 2, 7, 10, 12 \\
~57	& J1023+4334	& 10 23 05.05	& +43 34 33.2	& EE	& ---  	& ~145	& ~~816	& --0.77&---& ---	& ---	 	\\
~58	& J1031+5225	& 10 31 43.51	& +52 25 35.1	& SDSS	& 0.17 	& ~920	& ~4846	& --0.74&~275& ~10.77& 1, 2, 4, 10 \\
~~~59$^a$& J1032+2730 	& 10 32 57.60	& +27 30 15.1	& SDSS	& ---  	& ~~72	& ~~262	& --0.58&---& ---	& 1, 16 \\
\hline
\end{tabular}
\end{centering}
\end{footnotesize}
\end{table*}

\begin{table*}
\contcaption{\bf Candidate {\it X}-Shaped Radio Sources}
\begin{footnotesize}
\begin{centering}
\begin{tabular}{cccccccccccl}
\hline
Catalog &  Name         & R.A.          & Decl.         & Ref.     &Redshift&$F_{1400}$&$F_{150}$&$\alpha_{150}^{1400}$&Linear&$L$&Other Catalogs \\
Number  &               & (J2000.0)     & (J2000.0)     &    & ($z$)    & (mJy)    & (mJy)        &            &Size& (ergs$^{-1}$)  &    \\
	&               &               &               &       &        &          &         &     &(kpc)&             ($\times 10^{42}$) &  \\
% (1)   &   (2)         & (3)           &  (4)          & (5)    &  (6)   & (7)      & (8)        & (9)        & (10)        & (11)          & (12)    \\
\hline
~60	& J1049+5711	& 10 49 06.22	& +57 11 53.0	& EE	& ---  	& ~~16 	& ~~~54	& --0.54&---& ---	& 1, 7	\\
~61	& J1050+3240	& 10 50 11.17	& +32 40 24.8	& SDSS	& ---  	& ~~33	& ~~119	& --0.56&---& ---	& 1, 7  \\
~62	& J1107+5716    & 11 07 38.59   & +57 16 00.4   & EE   	& ---  	& ~~13 	& ---	& ---	&---& ---	& 1          \\
~63	& J1115+4314    & 11 15 21.25   & +43 14 37.8 	& SDSS 	& 0.46 	& ~350	& ~1611	& --0.68&~540& ~42.78& 6		\\
~64	& J1122+0046	& 11 22 00.98	& +00 46 31.5	& EE	& ---  	& ~293	& ~~910	& --0.51&---& ---	& 1, 8, 10 \\
~65	& J1124+4325	& 11 24 12.23	& +43 25 50.0	& SDSS	& 2.24 	& ~168	& ~~698	& --0.64&3347& 775.75& 1, 2, 6, 10, 15 \\
~66	& J1124+1717	& 11 24 57.40	& +17 17 44.7	& SDSS	& 0.14 	& ~~12	& ---	& ---	&~113& ---	& 1		\\
~67	& J1130+3434	& 11 30 05.74	& +34 34 28.8	& EE	& ---  	& ~205	& ~1315	& --0.83&---& ---	& 1, 7, 20	\\
~68	& J1134+3835	& 11 34 03.87	& +38 35 52.4	& SDSS	& 0.50 	& ~324	& ---	& ---	&~852& ---	& 1, 2, 12, 20 \\
~69	& J1136--0329	& 11 36 01.40	& --03 29 09.5  & EE	& --- 	& ~292	& ---	& ---	&---& ---	& 1, 2, 10 \\
~70	& J1138+1845	& 11 38 52.17	& +18 45 33.2	& SDSS	& 0.18 	& ~~22	& ~~~74	& --0.54&~172& ~~0.43& 1		\\
~71	& J1141--0723	& 11 41 47.83	& --07 23 24.7	& APMUKS& ---  	& ~114	& ---	& ---	&---& ---	& 1		\\
~72	& J1142+5832	& 11 42 23.84 	& +58 32 01.5	& EE	& ---  	& ~160	& ---	& ---	&---& ---	& ---		\\
~73	& J1144+1031	& 11 44 20.33	& +10 31 35.3	& SDSS	& 0.28 	& ~~38	& ~~160	& --0.64&~274& ~~1.58& 1		\\
~74	& J1149+2554	& 11 49 08.01	& +25 54 38.7	& EE	& ---  	& ~~62	& ~~410	& --0.84&---& ---	& ---		\\
~75	& J1149+4618	& 11 49 50.67	& +46 18 50.6	& SDSS	& 0.62 	& ~129	& ~~776	& --0.80&~1025& ~28.40& 1, 2, 6, 12, 21 \\
~76	& J1150+3622	& 11 50 50.16	& +36 22 03.7	& SDSS	& 0.14 	& ~540	& ~3435	& --0.83&~209& ~~4.16& 1, 2, 4, 10, 12, 20 \\
~77	& J1155--0646	& 11 55 52.78	& --06 46 37.9	& EE	& ---  	& ~~36	& ---	& ---	&---& ---	& 1		\\
~78	& J1157+0845	& 11 57 54.33	& +08 45 01.4	& SDSS	& 0.43 	& ~176	& ~~873	& --0.72&~754& ~17.58	& 1, 2, 10 \\
~79	& J1201+2520	& 12 01 24.72	& +25 20 23.7	& SDSS	& 0.50 	& ~857	& ~4160	& --0.71&~977& 123.97& 1, 2, 4, 20 \\
~80	& J1210+3157	& 12 10 37.57	& +31 57 06.0	& SDSS	& 0.39 	& ~274	& ---	& ---	&1019& ---	& 2, 7 \\
~81	& J1211+5717	& 12 11 22.96	& +57 17 51.7	& SDSS	& --- 	& ~220	& ~1999	& --0.99&---& ---	& 1, 2, 6 \\
~82	& J1213+1343	& 12 13 06.68	& +13 43 17.8	& SDSS	& 0.17 	& ~225	& ~~884	& --0.61&~229& ~~3.56	& 13, 16 \\
~83	& J1220+6053	& 12 20 04.90	& +60 53 16.5	& SDSS	& 0.34 	& ~~32	& ~~100	& --0.51&~383& ~~2.74	& 1		\\
~84	& J1222--0638	& 12 22 48.91	& --06 38 01.7	& EE	& ---  	& ~~95	& ---	& ---	&---& ---	& 1, 2	\\
~85	& J1235+2120	& 12 35 26.66	& +21 20 34.7	& SDSS	& 0.42 	& 2913	& 21017	& --0.88&2117& 249.94& 2, 3, 9, 10, 11, \\
~86	& J1238+2752	& 12 38 26.08	& +27 52 06.5	& EE	& --- 	& ~~72	& ~~268	& --0.59&---& ---	& ---		\\
~87	& J1239+1706	& 12 39 04.34	& +17 06 26.2	& SDSS	& ---  	& ~~44	& ~~272	& --0.82&---& ---	& ---		\\
~88	& J1242+4244	& 12 42 37.96	& +42 44 03.1	& SDSS	& ---  	& ~277	& ~1216	& --0.66&---& ---	& 4, 1	\\
~89	& J1247+0807	& 12 47 55.91	& +08 07 33.9	& SDSS	& 0.43 	& ~~68	& ~~288	& --0.65&~527& ~~7.44	& 1		\\
~90	& J1252--0120	& 12 52 30.70	& --01 20 17.3	& SDSS	& 0.36 	& ~269	& ---	& ---	&~612& ---	& 1, 2, 8, 10 \\
~~~91$^a$& J1258+4435	& 12 58 00.87	& +44 35 28.7	& SDSS	& ---  	& ~960	& ---	& ---	&---& ---	& 15	\\
~92	& J1300+2215    & 13 00 49.09 	& +22 15 47.4 	& SDSS 	& 0.22 	& ~~47	& ~~154	& --0.53&~298& ~~1.48	& 1, 13 	\\
~93	& J1311--0234   & 13 11 34.09 	& --02 34 08.0	& EE	& ---  	& ~381	& ---	& ---	&---& ---	& ---  		\\
~94	& J1311+4101	& 13 11 44.32	& +41 01 58.3	& EE	& ---  	& ~420	& ---	& ---	&---& ---	& ---		\\
~95	& J1325+1535	& 13 25 43.66	& +15 35 52.1	& SDSS	& ---  	& ~~69	& ~~512	& --0.90&---& ---	& ---		\\
~96	& J1333+0219	& 13 33 45.13	& +02 19 12.0	& SDSS	& 1.23 	& ~200	& ~~930	& --0.69&2249& 230.13& 1, 2, 9 \\
~~~97$^a$& J1336--0730	& 13 36 06.55	& --07 30 25.5	& APMUKS& ---   & ~~70	& ---	& ---	&---& ---	& 1	\\
~98	& J1336+3626	& 13 36 27.73	& +36 26 27.3	& EE	& ---  	& ~102	& ~~526	& --0.73&---& ---	& 1	\\
~99	& J1338+3851	& 13 38 49.80	& +38 51 11.7	& EE	& 0.25 	& 3461	& 23868	& --0.86&~363& ~87.07& 1, 2, 3, 10, 12, 20 \\
100	& J1349+1858	& 13 49 06.04	& +18 58 28.0	& SDSS	& 0.19 	& ~189	& ~~839	& --0.67&~279& ~~3.22& 1, 2, 10, 12 \\
101	& J1354+5840	& 13 54 45.99	& +58 40 00.7	& SDSS 	&5.7E--3& ~~15	& ~~~44	& --0.48&~~~3&3.19E--4& 1, 14, 15, 19 \\
~~102$^a$& J1358+1203	& 13 58 59.01	& +12 03 58.7	& EE	& ---  	& ~~43	& ~~184	& --0.65&---& ---	& 1		\\
103	& J1400+2719	& 14 00 08.00	& +27 19 14.1	& EE	& 0.16 	& ~534	& ~2738	& --0.73&~320& ~~5.51& 1, 2, 4, 20	\\
~~104$^a$& J1403+3827	& 14 03 11.75	& +38 27 59.4	& SDSS	& 0.54 	& ~674	& ~3419	& --0.73&1437& 112.94& 4		\\
105	& J1407+5131	& 14 07 24.20	& +51 31 24.4	& EE	& --- 	& ~420	& ---	& ---	&---& ---	& ---  		\\
106	& J1412+1739	& 14 12 17.00	& +17 39 48.0	& EE	& --- 	& ~~23	& ---	& ---	&---& ---	& 1		\\
107	& J1412+5839	& 14 12 47.81	& +58 39 34.7	& EE	& ---  	& ~~40	& ~~235	& --0.79&---& ---	& 1		\\
108	& J1417+4051	& 14 17 58.37	& +40 51 53.3	& SDSS	& ---  	& ~252	& ~1487	& --0.79&---& ---	& ---		\\
109	& J1418+2323	& 14 18 32.78	& +23 23 34.7	& EE	& ---  	& ~~22	& ~~110	& --0.72&---& ---	& ---		\\
110	& J1419+2303	& 14 19 42.59	& +23 03 27.6	& EE	& ---  	& ~~24	& ~~208	& --0.97&---& ---	& ---		\\
111	& J1420+5122	& 14 20 29.66	& +51 22 33.9	& SDSS	& 0.70 	& ~148	& ~~773	& --0.74&~771& ~44.98& 16	\\
112	& J1422+5022	& 14 22 25.25	& +50 22 13.3	& SDSS	& 0.18 	& ~~22	& ~~~69	& --0.51&~131& ~~0.47& 1, 14	\\
113	& J1422+1935	& 14 22 58.76	& +19 35 45.8	& EE	& ---  	& 3659	& ---	& ---	&---& ---	& ---		\\
~~114$^a$& J1424+0025	& 14 24 19.98	& +00 25 36.5	& SDSS	& ---  	& ~159	& ~~714	& --0.67&---& ---	& 1, 2, 8, 10	\\
~~115$^a$& J1440+6035	& 14 40 48.22	& +60 35 46.0	& SDSS	& ---  	& ~166	& ~~849	& --0.73&---& ---	& 1, 6	\\
116	& J1443+1440	& 14 43 00.08	& +14 40 17.0	& EE	& ---  	& ~~66	& ~~276	& --0.64&---& ---	& 12		\\
~~117$^a$& J1454+0959	& 14 54 05.82	& +09 59 52.6	& SDSS	& ---  	& ~100	& ~~329	& --0.53&---& ---	& 12 		\\
~~118$^a$& J1504+5749	& 15 04 08.06	& +57 49 22.6	& SDSS	& ---  	& ~~94	& ---	& ---	&---& ---	& 1		\\
~~119$^a$& J1506+2027	& 15 06 19.14	& +20 27 40.8	& SDSS	& ---  	& ~293	& ~2353	& --0.93&---& ---	& 12		\\
120	& J1516+3205	& 15 16 47.11	& +32 05 14.7	& SDSS	& 0.54 	& ~112	& ~~389	& --0.56&~480& ~23.56 & 1, 2, 6, 12	\\
121	& J1519+5342	& 15 19 36.72	& +53 42 55.4	& SDSS	& 0.48 	& ~479	& ---	& ---	&~772& ---	& 15	\\
\hline
\end{tabular}
\end{centering}
\end{footnotesize}
\end{table*}

\begin{table*}
\contcaption{\bf Candidate {\it X}-Shaped Radio Sources}
\begin{footnotesize}
\begin{centering}
\begin{tabular}{cccccccccccl}
\hline
Catalog &  Name         & R.A.          & Decl.         & Ref.     &Redshift&$F_{1400}$&$F_{150}$&$\alpha_{150}^{1400}$&Linear&$L$&Other Catalogs \\
Number  &               & (J2000.0)     & (J2000.0)     &    & ($z$)    & (mJy)    & (mJy)        &            &Size& (ergs$^{-1}$)  &    \\
	&               &               &               &       &        &          &         &     &(kpc)&               ($\times 10^{42}$) &  \\
% (1)   &   (2)         & (3)           &  (4)          & (5)    &  (6)   & (7)      & (8)        & (9)        & (10)        & (11)          & (12)    \\
\hline
122	& J1519--0408	& 15 19 49.79	& --04 08 42.9	& EE	& ---  	& ~231	& ---	& ---	&---& ---	& 2, 8, 10 \\
123	& J1523+1130	& 15 23 27.56	& +11 30 23.9	& SDSS	& 0.20 	& ~413	& ~1258	& --0.50&~247& ~11.37 & 1, 2, 12, 15 \\
~~124$^a$& J1523+2116	& 15 23 31.75	& +21 16 56.8	& SDSS	& ---  	& ~~60	& ~~223	& --0.59&---& ---	& 1		\\
125	& J1537+3902	& 15 37 49.51	& +39 02 37.6	& SDSS	& ---  	& ~~94	& ~~492	& --0.74&---& ---	& 21		\\
126	& J1539+5030	& 15 39 56.56	& +50 30 08.7	& EE	& ---  	& ~~85	& ~~279	& --0.53&---& ---	& 1		\\
127	& J1553+2811	& 15 53 07.01	& +28 11 24.7	& EE	& ---  	& ~~54	& ~~226	& --0.64&---& ---	& 1		\\
128	& J1553+2348	& 15 53 43.59	& +23 48 25.4	& SDSS	& 0.12 	& ~618	& ---	& ---	&~424& ---	& 2, 4, 11, 12, 14, 15 \\
129	& J1558+3404	& 15 58 31.84	& +34 04 44.0	& SDSS	& 0.51 	& ~~87	& ~~329	& --0.60&~556& ~15.17& 1		\\
130	& J1605+1743	& 16 05 08.98	& +17 43 47.6	& SDSS	& 0.03 	& ~793	& ~3092	& --0.61&~~40& ~~0.36	& 1, 2, 4, 11, 14, 19 \\
131	& J1607+1551	& 16 07 06.99	& +15 51 33.9	& SDSS	& 0.50 	& ~651	& ~2044	& --0.51&~669& 123.79& ---		\\
132	& J1617+3222	& 16 17 42.53	& +32 22 34.3	& SDSS	& 0.15 	& 2587	& 10002	& --0.60&~386& ~30.54 & 2, 3, 10, 11, 20 \\
133	& J1620+1736	& 16 20 21.82	& +17 36 24.0	& SDSS	& 0.56 	& 1995	& 16503	& --0.94&1148& 330.25& 2, 4, 11, 12, 15 \\
134	& J1626+4827	& 16 26 54.30	& +48 27 39.1	& SDSS	& ---  	& ~~37	& ~~167	& --0.67&---& ---	& 1		\\
135	& J1628+1658	& 16 28 31.47	& +16 58 33.3	& SDSS	& ---  	& ~~38	& ~~160	& --0.64&---& ---	& 1		\\
136	& J1633+3025	& 16 33 23.72	& +30 25 00.4	& SDSS	& 0.57 	& ~124	& ~~482	& --0.61&~565& ~26.93 & 20		\\
~~137$^a$& J1635+3722	& 16 35 02.53	& +37 22 14.7	& SDSS	& 0.46 	& ~645	& ~2630	& --0.63&~862& ~84.95 & 4		\\
138	& J1643+3729	& 16 43 06.16	& +37 29 30.3	& SDSS	& 0.56	& 1440	& ~8535	& --0.80&~906& 246.99& ---		\\
139	& J1650+3455	& 16 50 26.81	& +34 55 36.0	& SDSS	& 0.19 	& ~216	& ~~788	& --0.58&~255& ~~4.42 & 20		\\
140	& J1708+2435	& 17 08 46.13	& +24 35 28.8	& SDSS	& 1.36 	& ~369	& ~2176	& --0.79&2890& 508.52& ---	\\
141 & J1709+3425 & 17 09 39.15 & +34 25 50.8 & SDSS & 0.08  & ~646 & ~3534 & --0.76&~113& ~~1.58 & 4, 10, 11, 12 \\
142	& J1712+3857	& 17 12 11.13	& +38 57 06.8	& SDSS	& ---  	& ~192	& ~~724	& --0.59&---& ---	& 1		\\
~~143$^a$& J1715+5420	& 17 15 39.44	& +54 20 59.6	& SDSS	& 0.18 	& ~320	& ~2642	& --0.94&~229& ~~4.24	& 1, 2, 4, 10, 12 \\
144	& J1719+6155	& 17 19 25.16	& +61 55 33.9	& EE	& ---  	& ~360	& ~1989	& --0.76&---& ---	& ---		\\
145	& J1742+5917	& 17 42 43.67	& +59 17 07.3	& SDSS	& ---  	& 1090	& ~6030	& --0.76&---& ---	& ---		\\
146	& J1742+6145	& 17 42 51.59	& +61 45 54.6	& SDSS	& 0.54 	& 1347	& ~7048	& --0.74&~897& 216.66& 6		\\
147	& J2122+0001	& 21 22 17.10	& +00 01 15.5	& SDSS	& 0.42 	& ~157	& ~~722	& --0.68&~491& ~15.73	& 1, 12 \\
148	& J2123+1033    & 21 23 19.52 	& +10 33 26.0   & SDSS 	& ---  	& ~154 	& ~~576	& --0.59&---& ---	& 1  	\\
149	& J2153+0025	& 21 53 24.83	& +00 25 25.7	& EE	& ---  	& ~~38	& ~~128	& --0.54&---& ---	& 1		\\
~~150$^a$& J2213--0854	& 22 13 12.57	& --08 54 34.1	& EE	& ---  	& ~600	& ---	& ---	&---& ---	& ---		\\
151	& J2213--0544	& 22 13 35.87	& --05 44 22.7	& EE	& ---  	& ~~32	& ---	& ---	&---& ---	& 1		\\
152	& J2215--0525	& 22 15 01.79	& --05 25 17.5	& 2MASX	& 0.09$^{33}$ 	& ~216	& ---	& ---	&~161& ---	& ---		\\
153	& J2218+0012	& 22 18 30.16	& +00 12 21.2	& SDSS	& 0.29 	& ~412$^\ast$	& ~2763	& --0.85&~627& ~15.44 & ---		\\
154	& J2221--0326	& 22 21 29.54	& --03 26 16.8	& EE	& 0.36 	& ~~83	& ---	& ---	&~475& ---	& 1		\\
~~155$^a$& J2243--0954	& 22 43 49.64	& --09 54 07.7	& SDSS	& ---  	& ~~87	& ---	& ---	&---& ---	& ---		\\
156	& J2248--0449   & 22 48 30.44 	& --04 49 45.2  & EE	& ---  	& ~~43 	& ---	& ---	&---& ---	& 1  	\\
157	& J2249+0209	& 22 49 40.32	& +02 09 28.4	& EE   	& ---  	& ~182	& ~1894	& --1.05&---& ---	& 1, 2, 12\\
158	& J2257--0603	& 22 57 50.82	& --06 03 42.5	& EE	& ---  	& ~~46	& ---	& ---	&---& ---	& 1		\\
159	& J2301+1136	& 23 01 57.79	& +11 36 46.2	& EE	& ---  	& ~161	& ~~860	& --0.75&---& ---	& 1, 2, 10, 12\\
160	& J2324+1438	& 23 24 32.09	& +14 38 21.9	& SDSS	& 0.04 	& ~116$^\ast$	& ---	& ---	&~~46& ---	& 1, 2, 11, 29 \\
161	& J2351--0109	& 23 51 56.12	& --01 09 13.4	& SDSS	& 0.17 	& 1624	& ---	& ---	&~223& ---	& 1, 2, 4, 8, 9, 10, 11, 14 \\
\hline
\end{tabular}
\end{centering}
References--  
	1: NVSS \citep{Co98}; 2: VLA Low-Frequency Sky Survey (VLSS) \citep{Co07}; 3: 3C \citep{Be62, Ed59}; 4: 4C \citep{Pi65, Go67, Ca69}; 5: 5C \citep{Ke66, Po68, Po69, Wi70, Pe75, Wa77, Pe78, Sc81, Be82, Be88}; 6: 6C \citep{Ba85, Ha88, Ha90, Ha91, Ha93a, Ha93b}; 7: 7C \citep{Mc90, Ko94, Wa96, Ve98}; 8: Parkes-MIT-NRAO (PMN) \citep{Gr94}; 9: The Parkes catalogue of radio sources (PKS) \citep{Bo64}; 10: Texas Survey of Radio Sources (TXS) \citep{Do96}; 11: Cul \citep{Sl95}; 12: 87GB \citep{Gr91}; 13: Automatic Spectroscopic K-means-based classification (ASK) \citep{Sa11}; 14: 2 Micron All Sky Survey Extended objects - Final Release (2MASX) \citep{Sk06}; 15: GALaxy Evolution eXplorer all-sky catalog (GALEXASC) \citep{Ag05}; 16: Galaxy Evolution Explorer Medium-deep Sky Catalog (GALEXMSC) \citep{Ag05}; 17: Gaussian Mixture Brightest Cluster Galaxy (GMBCG) \citep{Ha10}; 18: Maximum likelihood redshift Brightest Cluster Galaxy (MaxBCG) \citep{Ko07}; 19: New General Catalogue (NGC) \citep{Dr88}; 20: B2 \citep{Co70, Co72, Co73, Fa74b}; 21: B3 \citep{Fi85}; 22: RC \citep{Go92}; 23: WHL \citep{We10}; 24: MGC \citep{Vo62}; 25: Abell Clusters of Galaxies (ABELL) \citep{Ab58, Ab89}; 26: Wide-field Infrared Survey Explorer (WISE) \citep{Ch11, Re11}; 27: WISE Blazar-like Radio-loud Sources (WB) \citep{Wh92}; 28: Automated Plate Measurement United Kingdom Schmidt (APMUKS) \citep{Ma90}; 29: Galaxy Identification Number (GIN) \citep{We96}; 30: \citet{Ho09}; 31: \citet{He91}; 32: \citet{Br05}; 33: \citet{Jo09}; Spitzer Space Telescope Source List - version 4.2 (SSTSL2) \citep{Ca13}; Las Campanas Redshift Survey (LCRS) \citep{Sh96}; Texas--Oxford NVSS Structure $08^h$ region (TONS08) \citep{Br03}; Texas--Oxford NVSS Structure $08^h$ region wider survey(TONS08w) \citep{Br05}

Notes--- 
$^a$ The soucre is present in \cite{Pr11}.\\
$\ast$ The FIRST flux (at 1400 MHz) is used instead of the NVSS flux (at 1400 MHz).
\end{footnotesize}
\end{table*}

\begin{table*}
\caption{\bf Candidate {\it S/Z}-Shaped Radio Sources}
\begin{footnotesize}
\begin{centering}
\begin{tabular}{cccccccccccl}
\hline
Catalog &  Name         & R.A.          & Decl.         & Ref.     &Redshift&$F_{1400}$&$F_{150}$&$\alpha_{150}^{1400}$&Linear&$L$&Other Catalogs \\
Number  &               & (J2000.0)     & (J2000.0)     &    & ($z$)    & (mJy)    & (mJy)        &            &Size& (ergs$^{-1}$)  &    \\
	&               &               &               &       &        &          &         &     &(kpc)&               ($\times 10^{42}$) &  \\
% (1)   &   (2)         & (3)           &  (4)          & (5)    &  (6)   & (7)      & (8)        & (9)        & (10)        & (11)          & (12)    \\
\hline
~~1	& J0002--0411	& 00 02 41.20	& --04 11 55.2	& GALEXASC& --- & ~216	& ---	& ---	&---& ---	& 1, 2, 8, 10	\\
~~2	& J0047+1221	& 00 47 11.67	& +12 21 24.9	& EE	& ---  	& ~~34	& ~~140	& --0.63&---& ---	& ---		\\
~~3	& J0126--0118	& 01 26 04.13	& --01 18 22.2	& EE	& ---  	& ~389$^\ast$& ---& ---	&---& ---	& ---		\\
~~4	& J0133+0957	& 01 33 16.94	& +09 57 30.8	& SDSS	& 0.19 	& ~~27	& ~~~72	& --0.44&~192& ~~0.76	& 1, 14	\\
~~5	& J0141+1213	& 01 41 09.94	& +12 13 52.0	& 2MASX	& ---  	& ~348	& ---	& ---	&---& ---	& ---		\\
~~6	& J0145--0820	& 01 45 56.65	& --08 20 26.3	& SDSS	& 0.19 	& ~~42	& ---	& ---	&~254& ---	& 1		\\
~~7	& J0204--0915	& 02 04 06.30	& --09 15 59.8	& EE	& ---  	& ~207	& ---	& ---	&---& ---	& 1, 2, 8, 10 \\
~~8	& J0204--0716	& 02 04 54.96 	& --07 16 03.3	& EE	& ---  	& ~263	& ---	& ---	&---& ---	& 1, 2, 8, 10 \\
~~~~9$^a$& J0206--0215	& 02 06 13.50	& --02 15 04.1	& EE	& ---  	& ~~24	& ---	& ---	&---& ---	& 1		\\
~10	& J0210--0310	& 02 10 30.61	& --03 10 47.0	& SDSS	& 0.24 	& ~~25	& ---	& ---	&~240& ---	& 1	\\
~~~11$^b$& J0219+0155	& 02 19 58.72	& +01 55 48.9	& 2MASS	& 0.04$^{30}$ 	& ~548	& ~2063	& --0.59&~~79& ~~0.45	& 1, 2, 8, 9, 10 \\
~12	& J0252--0756	& 02 52 27.53	& --07 56 05.4	& SDSS	& 0.08 	& ~112	& ---	& ---	&~~91& ---	& 1, 2, 13, 14 \\
~13	& J0256+0016    & 02 56 40.12 	& +00 16 33.5  	& SDSS 	& ---  	& ~~20	& ~~~92	& --0.68&---& ---	& 1      	\\
~14	& J0313--0631  	& 03 13 29.05   & --06 31 19.5  & EE   	& ---  	& ~~80	& ---	& ---	&---& ---	& 1      	\\
~15	& J0325--0743	& 03 25 23.18	& --07 43 48.5	& SDSS	& --- 	& ~~14	& ---	& ---	&---& ---	& 1		\\
~~~16$^b$& J0710+3546	& 07 10 31.01	& +35 46 50.5	& EE	& ---  	& ~~76	& ---	& ---	&---& ---	& 1, 6, 12 \\
~17	& J0727+4228	& 07 27 28.88	& +42 28 00.6	& SDSS	& ---  	& ~~54	& ~~175	& --0.53&---& ---	& 1, 16 \\
~~~18$^a$& J0728+4935	& 07 28 01.48	& +49 35 13.2	& 2MASX	& 0.08$^{31}$ 	& ~357	& ~1531	& --0.65&~264& ~~1.01	& 2, 10, 12 \\
~19	& J0729+4142	& 07 29 55.79	& +41 42 20.0	& EE	& --- 	& ~117	& ~~488	& --0.64&---& ---	& 1, 2, 12, 21	\\
~~~20$^a$& J0738+4820	& 07 38 37.70	& +48 20 47.9	& SDSS	& ---  	& ~~26	& ~~102	& --0.61&---& ---	& 1, 14	\\ 
~21	& J0738+3846	& 07 38 54.81	& +38 46 27.8	& SDSS	& 0.32 	& ~127	& ---	& ---	&~776& ---	& 2, 20	\\
~~~22$^b$& J0741+4618	& 07 41 01.91	& +46 18 41.6	& EE	& ---  	& ~230	& ~1228	& --0.75&---& ---	& 1, 2, 10, 12, 21  \\
~23	& J0746+4217	& 07 46 40.44	& +42 17 09.2	& SDSS	& 0.43 	& ~~14	& ---	& ---	&~491& ---	& 1 		\\
~24	& J0814+5059	& 08 14 31.27	& +50 59 23.1	& EE	& ---  	& ~~60	& ~~329	& --0.76&---& ---	& 1, 6	\\
~25	& J0816+0458	& 08 16 27.35 	& +04 58 47.6	& 2MASX	& ---  	& ~~61	& ~~~92	& --0.18&---& ---	& 8, 12, 22	 \\
~~~26$^a$& J0818+2247	& 08 18 54.09	& +22 47 44.9	& SDSS	& 0.10 	& ~308	& ~1113	& --0.58&~146& ~~1.48	& 1, 2, 10, 12 \\
~~~27$^a$& J0831+0125	& 08 31 52.79	& +01 25 53.3	& SDSS	& 0.27 	& ~125	& ~~844	& --0.86&~468& ~~3.88	& ---		\\
~28	& J0832+4559	& 08 32 58.03	& +45 59 28.2	& SDSS	& 0.64 	& ~~37	& ~~266	& --0.88&~710& ~~8.38	& 1		\\
~29	& J0839--0141	& 08 39 17.84	& --01 41 58.1	& MaxBCG& 0.28 	& ~~99	& ---	& ---	&~432& ---	& 1		\\
~30	& J0847+4934	& 08 47 51.14	& +49 34 14.1	& EE	& ---  	& ~234	& ~1712	& --0.89&---& ---	& 1, 2, 10, 12 \\
~31	& J0847+3147    & 08 47 58.65   & +31 47 50.2 	& EE   	& ---  	& ~164$^\ast$	& ---	& ---	&---& ---	& ---           \\
~32	& J0850+0153	& 08 50 51.71	& +01 53 09.8	& EE	& ---  	& ~~50	& ~~192	& --0.60&---& ---	& ---		\\
~33	& J0850+4753	& 08 50 53.41	& +47 53 51.6	& SDSS	& 0.18	& ~~67	& ~~273	& --0.63&~219& ~~1.17	& 1, 2, 14 \\
~34	& J0856+4951	& 08 56 18.72	& +49 51 06.4	& SDSS	& 0.62 	& ~~37	& ~~167	& --0.67&~740& ~~9.15	& 1		\\
~~~35$^b$& J0859--0252	& 08 59 54.15	& --02 52 41.8	& EE	& ---  	& ~402	& ---	& ---	&---& ---	& 1, 2, 4, 8, 9, 10 \\
~~~36$^a$& J0902+5203	& 09 02 36.84	& +52 03 48.7	& SDSS	& 0.10 	& ~425	& ~2060	& --0.71&~229& ~~1.75	& 1, 2, 6, 10, 12, 14  \\
~37	& J0906+4752	& 09 06 20.55	& +47 52 08.1	& SDSS	& 0.24 	& ~154	& ---	& ---	&~446& ---	& 14		\\
~~~38$^a$& J0918--0650	& 09 18 13.05	& --06 50 59.1	& EE	& ---  	& ~~58	& ---	& ---	&---& ---	& 1		\\
~~~~~~39$^{a,b}$& J0924+4034& 09 24 01.17& +40 34 57.1	& SDSS	& 0.16 	& ~320	& ~2022	& --0.82&~313& ~~3.21	& 1, 2, 10, 12, 14, 15, 21 \\
~40	& J0926+2921	& 09 26 38.99	& +29 21 22.6	& SDSS	& --- 	& ~126	& ---	& ---	&---& ---	& ---		\\
~~~41$^a$& J0933--0310	& 09 33 10.69	& --03 10 40.1	& EE	& --- 	& ~~75	& ---	& ---	&---& ---	& 1		\\
~42	& J0940+1510    & 09 40 13.55   & +15 10 53.2 	& SDSS  & 0.63 	& ~~26	& ~~104	& --0.62&~685& ~~6.98	& 1 	   	\\
~43	& J0941+0814	& 09 41 42.11	& +08 14 0.5	& EE	& ---  	& ~~38	& ~~150	& --0.61&---& ---	& ---		\\
~44	& J1004--0337	& 10 04 23.49	& --03 37 26.8	& EE	& ---  	& ~~57	& ---	& ---	&---& ---	& 1		\\
~45	& J1006--0621	& 10 06 48.08	& --06 21 56.9	& EE	& ---  	& ~267	& ---	& ---	&---& ---	& 1, 2, 8, 10 \\
~~~~~~46$^{a,b}$& J1011--0607   & 10 11 34.79   & --06 07 53.0	& EE    & ---  	& ~~36	& ---	& ---	&---& ---	& 1		\\
~47	& J1011+4353    & 10 11 49.38   & +43 53 19.1	& SDSS  & ---  	& ~329	& ~1611	& --0.71&---& ---	& 1, 2, 6, 10, 12, 21 \\
~~~48$^a$& J1014+3507	& 10 14 18.28	& +35 07 29.1	& SDSS	& ---  	& ~157	& ~~937	& --0.80&---& ---	& 1, 2, 7, 10, 12, 20 \\
~~~49$^b$& J1018+1058	& 10 18 04.89	& +10 58 41.7	& EE	& ---  	& ~~44	& ~~186	& --0.64&---& ---	& 1		\\
~50	& J1028+0345	& 10 28 23.47	& +03 45 31.4	& SDSS	& 0.10 	& ~287	& ~1477	& --0.73&~279& ~~1.14	& 1, 2, 8, 10, 12, 14 \\
~~~51$^a$& J1028+1943	& 10 28 56.38   & +19 43 41.0	& SDSS	& 0.60 	& ~231 	& ~1584	& --0.86&~944& ~46.33	& 15	\\
~~~52$^a$& J1056+1128	& 10 56 14.77	& +11 28 42.7	& SDSS	& 0.42 	& ~~21	& ~~~92	& --0.66&~640& ~~2.11	& 1		\\
~53	& J1057+3012	& 10 57 20.44	& +30 12 30.4	& SDSS	& 0.24 	& ~~79	& ~~272	& --0.55&~396& ~~2.94	& 23		\\
~~~54$^a$& J1122+2125	& 11 22 29.39	& +21 25 02.4	& EE	& ---  	& ~630	& ---	& ---	&---& ---	& 1, 2, 7, 9, 10, 11 \\
~55	& J1127+1909    & 11 27 58.48   & +19 09 27.3   & EE  	& ---   & ~206	& ---	& ---	&---& ---	& ---		\\
~~~56$^a$& J1138+2039	& 11 38 50.24	& +20 39 18.3	& SDSS	& 0.18 	& ~~74$^\ast$& ~~267& --0.58&~225& ~~1.35& 1, 13, 14 \\
~57	& J1140+1743	& 11 40 16.98	& +17 43 40.3	& SDSS	& 0.01 	& 1154	& ~3884	& --0.54&~~12& ~~0.07	& 1, 2, 4, 10, 11, 14, 15, 19 \\
~58	& J1145--0227	& 11 45 30.97	& --02 27 11.2	& SDSS	& 0.13 	& ~148	& ---	& ---	&~362& ---	& 8		\\
~59	& J1149+3802	& 11 49 50.18	& +38 02 37.0	& SDSS 	& 0.53 	& ~~62	& ~~250	& --0.62&~533& ~11.35	& 1		\\
~~~~~~60$^{a,b}$& J1156+2138& 11 56 45.30& +21 38 09.1	& SDSS	& 0.62 	& ~202	& ~1004	& --0.72&~819& ~47.04	& 1, 2, 7, 12 \\

\hline
\end{tabular}
\end{centering}
\end{footnotesize}
\end{table*}

\begin{table*}
\contcaption{\bf Candidate {\it S/Z}-Shaped Radio Sources}
\begin{footnotesize}
\begin{centering}
\begin{tabular}{cccccccccccl}
\hline
	Catalog &  Name         & R.A.          & Decl.         & Ref.     &Redshift&$F_{1400}$&$F_{150}$&$\alpha_{150}^{1400}$&Linear&$L$&Other Catalogs \\
	Number  &               & (J2000.0)     & (J2000.0)     &    & ($z$)    & (mJy)    & (mJy)        &            &Size& (ergs$^{-1}$)  &    \\
	&               &               &               &       &        &          &         &     &(kpc)&               ($\times 10^{42}$) &  \\
% (1)   &   (2)         & (3)           &  (4)          & (5)    &  (6)   & (7)      & (8)        & (9)        & (10)        & (11)          & (12)    \\
\hline
~61	& J1157+3012	& 11 57 08.30	& +30 12 17.1	& EE	& ---  	& ~~34	& ~~231	& --0.86&---& ---	& 1		\\
~~~62$^b$& J1201+3257   & 12 01 51.87   & +32 57 01.3   & EE    & ---  	& ~158	& ---	& ---	&---& ---	& 6            \\
~~~63$^a$& J1204+5531	& 12 04 06.72	& +55 31 14.6	& SDSS	& --- 	& ~459	& ---	& ---	&---& ---	& 15	\\
~~~64$^a$& J1204+0345	& 12 04 25.76	& +03 45 05.4	& EE	& ---  	& ~~62	& ~~220	& --0.57&---& ---	& 1		\\
~65	& J1206--0621	& 12 06 26.31	& --06 21 43.4	& LCRS	& --- 	& ~130	& ---	& ---	&---& ---	& 1		\\
~66	& J1208+2513    & 12 08 09.07   & +25 13 57.0 	& EE	& ---  	& ~129$^\ast$	& ~~257	& --0.30	&---& ---	& 1  	\\
~67	& J1215+1709	& 12 15 43.82	& +17 09 17.6	& SDSS	& 0.10 	& ~488	& ~1462	& --0.49&~233& ~~2.81	& 1, 2, 10, 15 \\
~68	& J1218+3548    & 12 18 19.04   & +35 48 17.2 	& SDSS  & 0.25 	& ~129	& ~~576	& --0.67&~592& ~~4.08	& 7, 10, 12, 14, 17, 20 \\
~~~69$^a$& J1222+3758	& 12 22 09.68	& +37 58 55.7	& EE 	& ---  	& ~170	& ---	& ---	&---& ---	& ---		\\
~70	& J1223+2542	& 12 23 00.24	& +25 42 02.9	& SDSS	& 0.33	& ~~23	& ~~~80	& --0.56&~493& ~~1.58	& 1, 14	\\
~71	& J1232+3130	& 12 32 11.44	& +31 30 58.1	& SDSS	& 0.35 	& ~~80	& ~~320	& --0.62&~533& ~~5.82	& 1 		\\
~72	& J1234--0804   & 12 34 37.40   & --08 04 14.0  & EE    & ---  	& ~~88	& ---	& ---	&---& ---	& 1  	\\
~~~73$^a$& J1243--0613	& 12 43 34.68	& --06 13 22.2	& 2MASX	& 0.14$^{32}$ 	& ~367	& ---	& ---	&~357& ---	& 1, 2, 4, 8, 9, 10, 14, 15 \\
~~~74$^a$& J1255+4405	& 12 55 54.60	& +44 05 21.8	& SDSS	& ---  	& ~~51	& ~~156	& --0.50&---& ---	& ---		\\
~75	& J1300--0337	& 13 00 31.44	& --03 37 46.2	& SDSS	& ---  	& ~170	& ---	& ---	&---& ---	& ---		\\
~~~76$^a$& J1303+0339  	& 13 03 59.47 	& +03 39 32.3 	& SDSS  & 0.18 	& ~420 	& ~2530	& --0.49 &~657& ~~5.77	& 2, 4, 9, 10 \\
~~~77$^a$& J1304+2015  	& 13 04 22.28 	& +20 15 56.3 	& SDSS  & 0.19 	& ~~66	& ~~331	& --0.72&~225& ~~1.09	& 1, 13, 14, 18, 23 \\
~78	& J1305+3622	& 13 05 51.80 	& +36 22 53.0	& MaxBCG& 0.31 	& ~211	& ~~910	& --0.65&~417& ~~9.85	& 20		\\
~~~79$^a$& J1307+1218	& 13 07 55.53	& +12 18 48.3	& SDSS	& 0.18 	& ~~60	& ~~199	& --0.54&~275& ~~1.13	& 1, 15 \\
~~~80$^a$& J1308--0500	& 13 08 39.03	& --05 00 37.5	& 2MASX	& ---  	& ~272	& ---	& ---	&---& ---	& 28	\\
~81	& J1310+2600	& 13 10 46.12	& +26 00 58.5	& SDSS	& --- 	& ~100	& ~~348	& --0.56&---& ---	& ---		\\
~~~82$^a$& J1319+0502	& 13 19 43.59	& +05 02 43.0   & SDSS 	& 1.28 	& ~~27	& ---	& ---	&2457& ---	& 1		\\
~~~83$^a$& J1320+2532	& 13 20 00.19	& +25 32 43.8	& SDSS	& 0.13 	& ~316	& ~1380	& --0.66&~207& ~~2.68	& 1, 2, 7, 14, 20 \\
~~~84$^a$& J1325+5736	& 13 25 11.19	& +57 36 01.2	& SDSS	& 0.12  & ~110	& ~1643	& --1.21&~297& ~~0.68	& 14, 24, 25 \\
~~~85$^a$& J1327+0007	& 13 27 57.49	& +00 07 51.3	& SDSS	& 0.51  & ~~62	& ~~275	& --0.67&~949& ~~9.61	& 1		\\
~86	& J1328+0150	& 13 28 42.37	& +01 50 59.5	& SDSS	& ---  	& ~~10	& ~~~43	& --0.65&---& ---	& 1		\\
~87	& J1337+5348	& 13 37 57.66	& +53 48 25.2	& SDSS	& 0.10 	& ~~29	& ~~~72	& --0.41&~121& ~~0.23	& 1		\\
~~~88$^a$& J1339+1024	& 13 39 13.64	& +10 24 47.9	& SDSS	& ---  	& ~116	& ~~782	& --0.85&---& ---	& 1		\\
~89	& J1345+3124	& 13 45 41.65	& +31 24 06.4	& SDSS	& 0.22	& ~~80	& ~~214	& --0.44&~258& ~~3.05	& 1, 14	\\
~~~90$^a$& J1346+6220	& 13 46 17.54	& +62 20 45.4	& SDSS 	& 0.12 	& ~160 	& ~~673	& --0.64&~173& ~~1.03	& 1, 2, 6, 10, 12, 15, 26, 27 \\
~~~91$^a$& J1349+4306	& 13 49 29.86	& +43 06 10.2	& EE	& ---  	& ~163	& ~1129	& --0.87&---& ---	& 21		\\
~92	& J1356+4555	& 13 56 59.94	& +45 55 04.6	& 2MASX	& 0.24 	& ~~95	& ~~355	& --0.59&~289& ~~3.22   & 1, 21	\\
~~~93$^a$& J1359+6119	& 13 59 54.22	& +61 19 45.2	& SDSS	& 0.18 	& ~~30	& ~~104	& --0.56&~210& ~~0.55	& 1	\\
~94	& J1402+4612	& 14 02 32.38	& +46 12 33.7	& SDSS	& 0.56 	& ~~93	& ~~500	& --0.75&~883& ~16.25   & 1, 2, 12, 15, 21 \\
~95	& J1404+3701	& 14 04 11.63	& +37 01 26.6	& EE	& ---  	& ~~26	& ~~142	& --0.76&---& ---	& 1		\\
~~~96$^a$& J1410+3014	& 14 10 50.88	& +30 14 09.6	& 2MASX	& 0.18$^{33}$ 	& ~~84	& ~~235	& --0.46&~338& ~~1.84	& 1, 18	\\
~97	& J1411+4535	& 14 11 10.93	& +45 35 18.1	& EE	& ---  	& ~~35	& ~~~86	& --0.40&---& ---	& 1		\\
~98	& J1417+0812	& 14 17 31.27	& +08 12 30.1	& SDSS	& 0.06 	& ~352	& ---	& ---	&~208& ---	& 2, 14, 15, 19 \\
~~~99$^a$& J1418+0952	& 14 18 13.47	& +09 52 38.5	& SDSS	& 0.34 	& ~333	& ~~692	& --0.33&~622& ~42.84	& 1, 2, 16\\
~~100$^a$& J1418+3530	& 14 18 40.50	& +35 30 01.6	& 2MASX	& ---  	& ~~90	& ~~194	& --0.34&---& ---	& 1		\\
~~101$^a$& J1426+2158	& 14 26 23.01	& +21 58 39.5	& SDSS	& 0.20 	& ~~81	& ---	& ---	&~359& --- 	& 1		\\
102	& J1426+3411	& 14 26 59.61	& +34 11 59.6	& SDSS	& 0.13 	& ~304	& ---	& ---	&~326& ---	& 15	\\
103	& J1437+1616	& 14 37 16.71	& +16 16 04.1	& SDSS	& 0.27 	& ~~38	& ---	& ---	&~493& ---	& 1		\\
104	& J1439+1106  	& 14 39 28.37 	& +11 06 02.7 	& EE	& ---   & ~~43 	& ~~115	& --0.44&---& ---	& 1 		\\
105	& J1442+5043	& 14 42 19.18	& +50 43 57.9	& SDSS	& 0.17 	& ~205	& ~~878	& --0.65&~326& ~~3.04	& 1, 17	\\
106	& J1524+1627	& 15 24 19.50	& +16 27 12.4	& SDSS	& 0.15 	& ~191	& ~~262	& --0.14&~247& ~~7.35	& 14		\\
107	& J1526+0053	& 15 26 42.05	& +00 53 28.8	& SDSS	& 0.12 	& ~236	& ~~612	& --0.43&~165& ~~2.35	& 1, 2, 8, 12 \\
108	& J1529--0629   & 15 29 19.98   & --06 29 21.5  & EE	& ---  	& ~109	& ---	& ---	&---& ---	& 1  	\\
109	& J1530+3301	& 15 30 22.23	& +33 01 19.5	& EE	& ---  	& ~~12	& ~~~41	& --0.55&---& ---	& 1		\\
110	& J1530--0703	& 15 30 58.88	& --07 03 32.4	& 2MASS	& ---  	& ~147	& ---	& ---	&---& ---	& ---		\\
111	& J1536+2357	& 15 36 38.05	& +23 57 06.1	& SDSS	& 0.51 	& ~~59	& ~~410	& --0.87&~767& ~~7.89	& 1		\\
112	& J1536--0453	& 15 36 55.69	& --04 53 50.0	& EE	& ---  	& ~~25	& ---	& ---	&---& ---	& 1		\\
~~113$^a$& J1540+1110  	& 15 40 13.80 	& +11 10 08.6 	& EE	& ---  	& ~~27	& ~~118	& --0.66&---& ---	& 1  	\\
~~114$^a$& J1553--0323	& 15 53 06.36	& --03 23 18.5	& EE	& ---  	& ~134	& ---	& ---	&---& ---	& ---		\\
115	& J1604+2355	& 16 04 56.66	& +23 55 57.6	& SDSS	& 0.03 	& ~661	& ---	& ---	&~122& ---	& 1, 2, 4, 14, 15, 19 \\
~~116$^a$& J1608+4309  	& 16 08 04.52 	& +43 09 48.5 	& SDSS 	& 0.08 	& ~~76 	& ~~224	& --0.48&~~96& ~~0.35	& 1, 13, 14 \\
117	& J1608+3505	& 16 08 29.68	& +35 05 53.8	& SDSS	& 0.35 	& ~~37	& ~~153	& --0.64&~506& ~~2.60	& 1		\\
118	& J1613+1921	& 16 13 35.02	& +19 21 05.8	& SDSS	& ---  	& ~195	& ~~803	& --0.63&---& ---	& ---		\\
119	& J1617+1420	& 16 17 52.73	& +14 20 17.2	& SDSS	& ---  	& ~~37	& ~~156	& --0.64&---& ---	& 1		\\

\hline
\end{tabular}
\end{centering}
\end{footnotesize}
\end{table*}

\begin{table*}
\contcaption{\bf Candidate {\it S/Z}-Shaped Radio Sources}
\begin{footnotesize}
\begin{centering}
\begin{tabular}{cccccccccccl}
\hline
	Catalog &  Name         & R.A.          & Decl.         & Ref.     &Redshift&$F_{1400}$&$F_{150}$&$\alpha_{150}^{1400}$&Linear&$L$&Other Catalogs \\
	Number  &               & (J2000.0)     & (J2000.0)     &    & ($z$)    & (mJy)    & (mJy)        &            &Size& (ergs$^{-1}$)  &    \\
	&               &               &               &       &        &          &         &     &(kpc)&               ($\times 10^{42}$) &  \\
% (1)   &   (2)         & (3)           &  (4)          & (5)    &  (6)   & (7)      & (8)        & (9)        & (10)        & (11)          & (12)    \\
\hline
120	& J1633+4220  	& 16 33 13.40 	& +42 20 31.3 	& SDSS 	& 0.12 	& ~110	& ~~179	& --0.22&~197& ~~2.00	& 1, 12, 14, 21 \\
~~121$^a$& J1643+2642	& 16 43 18.86	& +26 42 35.3	& EE	& ---  	& ~~68	& ~~330	& --0.71&---& ---	& 1		\\
122	& J1649+5358	& 16 49 34.80	& +53 58 16.0	& 2MASX	& ---  	& ~176	& ~~542	& --0.50&---& ---	& ---		\\
~~123$^a$& J1728+4200	& 17 28 43.50	& +42 00 06.5	& EE	& ---  	& ~~31	& ---	& ---	&---& ---	& ---		\\
124	& J2107--0203	& 21 07 44.25	& --02 03 40.4	& EE	& ---  	& ~~48	& ---	& ---	&---& ---	& 1		\\
125	& J2129--0549	& 21 29 00.88	& --05 49 52.9	& EE	& ---  	& ~~33	& ---	& ---	&---& ---	& 1		\\
126	& J2137--0811  	& 21 37 25.01 	& --08 11 05.6 	& SDSS 	& ---  	& ~~57 	& ---	& ---	&---& ---	& 1  	\\
127	& J2159--0211	& 21 59 15.62	& --02 11 25.3	& SDSS	& 0.30 	& ~~58	& ---	& ---	&~332& ---	& 1	\\
128	& J2210+1050	& 22 10 21.47	& +10 50 55.4	& EE	& ---  	& ~~63	& ~~265	& --0.64&---& ---	& 1, 2	\\
129	& J2250--0439	& 22 50 56.53	& --04 39 41.1	& EE	& ---  	& ~~30	& ---	& ---	&---& ---	& 1		\\
130	& J2306--0341	& 23 06 27.76	& --03 41 23.3	& EE	& ---  	& ~~51	& ---	& ---	&---& ---	& 1		\\
131	& J2307+1253	& 23 07 16.28	& +12 53 31.3	& SDSS	& 0.22 	& ~~57	& ~~206	& --0.57&~234& ~~1.67	& 1, 12	\\
132	& J2321--0912	& 23 21 21.90	& --09 12 42.8	& SDSS	& ---  	& ~~46	& ---	& ---	&---& ---	& 1		\\
~~133$^a$& J2322--0941 	& 23 22 08.20 	& --09 41 53.1 	& EE	& ---  	& ~125 	& ---	& ---	&---& ---	& 1  	\\
134	& J2331--0129	& 23 31 13.39	& --01 29 13.2	& EE	& --- 	& ~328	& ---	& ---	&---& ---	& 16	\\
~~135$^a$& J2339+0042  	& 23 39 00.34 	& +00 42 57.8 	& SDSS 	& 0.18 	& ~~29 	& ---	& ---	&~260& ---	& 1, 13, 14, 16\\
\hline
\end{tabular}
\end{centering}
References--  
	1: NVSS \citep{Co98}; 2: VLSS \citep{Co07}; 3: 3C \citep{Be62, Ed59}; 4: 4C \citep{Pi65, Go67, Ca69}; 5: 5C \citep{Ke66, Po68, Po69, Wi70, Pe75, Wa77, Pe78, Sc81, Be82, Be88}; 6: 6C \citep{Ba85, Ha88, Ha90, Ha91, Ha93a, Ha93b}; 7: 7C \citep{Mc90, Ko94, Wa96, Ve98}; 8: PMN \citep{Gr94}; 9: PKS \citep{Bo64}; 10: TXS \citep{Do96}; 11: Cul \citep{Sl95}; 12: 87GB \citep{Gr91}; 13: ASK \citep{Sa11}; 14: 2MASX \citep{Sk06}; 15: GALEXASC \citep{Ag05}; 16: GALEXMSC \citep{Ag05}; 17: GMBCG \citep{Ha10}; 18: MaxBCG \citep{Ko07}; 19: NGC \citep{Dr88}; 20: B2 \citep{Co70, Co72, Co73, Fa74b}; 21: B3 \citep{Fi85}; 22: RC \citep{Go92}; 23: WHL \citep{We10}; 24: MGC \citep{Vo62}; 25: ABELL \citep{Ab58, Ab89}; 26: WISE \citep{Ch11, Re11}; 27: WB \citep{Wh92}; 28: APMUKS \citep{Ma90}; 29: GIN \citep{We96}; 30: \citet{Fa99}; 31: \citet{Ma98}; 32: \citet{Jo09}; 33: \citet{Ry12}; SSTSL2 \citep{Ca13}; LCRS \citep{Sh96}; TONS08 \citep{Br03}; TONS08w \citep{Br05}

Notes-- 
$^a$ The soucre is present in \cite{Pr11}.\\
$^b$ The soucre is cataloged as an XRG by \cite{Ya19}.\\
%$^c$ the soucre is cataloged as XRG by \cite{Ya19} and also present in \cite{Pr11}.\\
$\ast$ The FIRST flux (at 1400 MHz) is used instead of the NVSS flux (at 1400 MHz).
\end{footnotesize}
\end{table*}

\end{document}